\documentclass[lettersize,journal]{IEEEtran}
\usepackage{amsmath,amsfonts,amssymb,amsthm}
\usepackage{algorithmic}
\usepackage{algorithm}
\usepackage{array}
\usepackage[caption=false,font=footnotesize,labelfont=rm,textfont=rm]{subfig}
\usepackage{textcomp}
\usepackage{stfloats}
\usepackage{url}
\usepackage{verbatim}
\usepackage{graphicx}
\usepackage{cite}
\usepackage{xcolor}
\begin{document}

\title{Energy Efficiency Maximization for Movable Antenna Communication Systems}

\author{Jingze Ding, \IEEEmembership{Graduate Student Member, IEEE},
		Zijian Zhou, \IEEEmembership{Member, IEEE}, 
		Lipeng Zhu, \IEEEmembership{Member, IEEE},\\
		Yuping Zhao, \IEEEmembership{Member, IEEE},
		Bingli Jiao, \IEEEmembership{Senior Member, IEEE},
		and Rui Zhang, \IEEEmembership{Fellow, IEEE}
\thanks{An earlier version of this paper was presented in part at the 2025 IEEE International Workshop on Radio Frequency and Antenna Technologies  (IEEE iWRF\&AT 2025) \cite{conference}.}
\thanks{This work was supported in part by the Project of General Administration of Customs of the People's Republic of China under Grant 2024HK290, in part by the Guangzhou Municipal Science and Technology Project under Grants 2023B04J0011 and 2025B04J0038, in part by The Guangdong Provincial Key Laboratory of Big Data Computing, in part by the National Natural Science Foundation of China under Grant 62331022, in part by the 2022 Stable Research Program of Higher Education of China under Grant 20220817144726001, and in part by the Guangdong Major Project of Basic and Applied Basic Research under Grant 2023B0303000001. \textit{ (Corresponding authors: Zijian Zhou; Rui Zhang.)}}
\thanks{Jingze Ding and Yuping Zhao are with the School of Electronics, Peking University, Beijing 100871, China (e-mail: djz@stu.pku.edu.cn; yuping.zhao@pku.edu.cn).}
\thanks{Zijian Zhou is with the School of Science and Engineering, The Chinese University of Hong Kong, Shenzhen 518172, China (e-mail: zijianzhou@link.cuhk.edu.cn).}
\thanks{Lipeng Zhu and Rui Zhang are with the Department of Electrical and Computer Engineering, National University of Singapore, Singapore 117583 (e-mail: zhulp@nus.edu.sg; elezhang@nus.edu.sg).}
\thanks{Bingli Jiao is with the School of Computing and Artificial Intelligence, Fuyao University of Science and Technology, Fuzhou 350109, China. He is also with the School of Electronics, Peking University, Beijing 100871, China (e-mail: jiaobl@pku.edu.cn)}
}
\maketitle

\begin{abstract}
This paper investigates energy efficiency maximization for movable antenna (MA)-aided multi-user uplink communication systems by considering the time delay and energy consumption incurred by practical antenna movement. We first examine the special case with a single user and propose an optimization algorithm based on the one-dimensional (1D) exhaustive search to maximize the user's energy efficiency. Moreover, we derive an upper bound on the energy efficiency and analyze the conditions required to achieve this performance bound under different numbers of channel paths. Then, for the general multi-user scenario, we propose an iterative algorithm to fairly maximize the minimum energy efficiency among all users. Simulation results demonstrate the effectiveness of the proposed scheme in improving energy efficiency compared to existing MA schemes that do not account for movement-related costs, as well as the conventional fixed-position antenna (FPA) scheme. In addition, the results show the robustness of the proposed scheme to imperfect channel state information (CSI) and provide valuable insights for practical system deployment.
\end{abstract}
\begin{IEEEkeywords}
Movable antenna (MA), energy efficiency, antenna position optimization, multi-user communication.
\end{IEEEkeywords}
\section{Introduction}
Next-generation wireless communication systems require increasingly higher capacity and more efficient resource allocation to support a wide range of modern applications \cite{nextgen}. In this context, multiple-input multiple-output (MIMO) technology \cite{mimo1} has become a cornerstone for enhancing the capacity and reliability of wireless networks. By deploying multiple spatially separated antennas at transceivers, MIMO systems can exploit the independent fading characteristics of multi-path channel components to achieve both spatial diversity and multiplexing gains, thereby significantly improving link reliability and spectral efficiency \cite{mimo2}. However, the antenna configurations implemented in existing systems are primarily based on the fixed-position antenna (FPA) \cite{mimo4}. While multiple FPAs can provide reliable performance via joint signal processing, they fall short in fully exploiting the degrees of freedom (DoFs) available in the continuous spatial domain, thus limiting the system's ability to adapt to varying channel conditions and user distributions.

To dynamically harness the continuous spatial DoFs in wireless channels, movable antenna (MA) technology \cite{MA1,add_MA1}, also known as fluid antenna with alternative implementation methods for antenna positioning \cite{FA}, has recently been proposed to overcome the inherent constraints of static antenna placement in conventional FPA systems. Besides, six-dimensional MA (6DMA) and other intelligent antenna systems \cite{6DMA1,6DMA2,6DMA4,polar, polar1} have also been proposed to improve the adaptability and reconfigurability of wireless communication systems. A typical implementation of MA is to connect each antenna to a radio frequency chain via a flexible cable, which allows its position to be adjusted in real time within a given spatial region using a stepper motor. Thanks to their flexibility, MAs can be strategically placed at positions with more favorable channel conditions to offer new capabilities in signal power improvement, interference mitigation, flexible beamforming, and spatial multiplexing \cite{MA1}. Given the aforementioned advantages, the MA technology has been widely applied to enhance the performance of existing wireless communication and sensing systems \cite{MA2,MA3,MA4,MA5,MAFD1,MAFD2,MAFD3,MAFD4,MAUAV1,MAUAV2,MAIRS1,MAIRS2,MAnear1,MAnear2,MAbeam,MAISAC1,MAISAC2}. The authors in \cite{MA2} first proposed the mechanical MA structure and developed the field-response channel model to describe the impact of MA movement on the phases of multiple channel paths. Then, the authors in \cite{MA3} extended the model to MA-aided point-to-point MIMO systems and showed that antenna position optimization can improve the MIMO channel capacity. Furthermore, the authors in \cite{MA4,MA5} studied multi-user uplink communication systems and demonstrated that flexible MA movement can reduce the correlation among user channel vectors, thereby helping to mitigate the multi-user interference effectively. Moreover, the authors in \cite{MAFD1,MAFD2,MAFD3,MAFD4} systematically investigated the performance improvements achieved through antenna position optimization in various full-duplex systems, including point-to-point full-duplex systems \cite{MAFD1}, secure full-duplex systems \cite{MAFD2,MAFD3}, and full-duplex satellite communication systems \cite{MAFD4}. In addition, the applications of MA arrays in unmanned aerial vehicle (UAV) communications \cite{MAUAV1,MAUAV2}, intelligent reflecting surface (IRS)-aided communications \cite{MAIRS1,MAIRS2}, near-field communications \cite{MAnear1,MAnear2}, flexible beamforming \cite{MAbeam}, and integrated sensing and communications (ISAC) \cite{MAISAC1,MAISAC2} were also explored.

The majority of existing MA studies concentrated on exploiting the MA movement to reconfigure wireless channels, thereby maximizing the achievable data rate or minimizing the transmit power. These works generally assumed that the time consumed by wavelength-scale MA movement is negligible. Nevertheless, in most practical scenarios, the MA moving delay cannot be ignored, as it reduces the time available for data transmission, thereby decreasing the achievable data rate. On the other hand, the operation of the stepper motor requires terminals equipped with MAs to consume additional energy for MA movement, which may sacrifice the energy available for data transmission in energy-constrained systems. Therefore, the movement-related time and energy consumptions are non-negligible and need to be considered in the design of MA systems. The authors in \cite{mini} first studied the MA moving delay and proposed an MA trajectory optimization algorithm. Besides, the authors in \cite{Th} optimized the communication throughput of the MA-aided multi-user downlink system within a given transmission block duration. Nevertheless, they did not consider the motor's energy consumption. Furthermore, the energy consumption of the stepper motor used for MA movement was investigated in \cite{energy1,energy2}. However, these works assumed a constant power consumption for the motor, which is only applicable to MA systems with fixed moving speeds and cannot reveal the impact of varying motor speeds on the system's performance. Generally, the power consumption of the motor increases with speed, but the increasing speed also reduces the MA moving delay. Therefore, it is necessary to develop a practical energy consumption model for stepper motors in MA systems that comprehensively accounts for the power consumption and time delay of MA movement.

In light of the above, this paper investigates the maximization of energy efficiency for MA-aided multi-user uplink communication systems by considering both the time delay and energy consumption incurred by MA movement, where each user is equipped with a single MA capable of linear movement. The main contributions of this paper are summarized as follows:
\begin{itemize}
	\item [1)] 
	First, we model the energy consumption of typical commercial stepper motors as a function of the MA's initial and optimized positions, based on which the energy efficiency for each user is defined for a given transmission block. To guarantee fairness in quality-of-service (QoS) for each user, we aim to maximize the minimum energy efficiency among all users by jointly optimizing their transmit power and MA positions, as well as the base station (BS)'s receive combining matrix, subject to the minimum communication throughput requirement and maximum transmit power for each user, and the finite moving region for each MA.
	\item [2)] 
	Next, for the special case of a single-user system, we propose a one-dimensional (1D) search algorithm to determine the optimal MA position and user's transmit power. To provide more insights, we derive the upper bound on the user's energy efficiency based on the optimized transmit power and analyze the conditions for achieving the performance bound under different numbers of channel paths.
	\item [3)] 
	Furthermore, for the general multi-user case, we decompose the original optimization problem into two sub-problems and propose an efficient algorithm for solving them based on successive convex approximation (SCA). By introducing slack variables, the proposed algorithm iteratively optimizes the receive combining matrix, transmit power, and MA positions until convergence is reached.
	\item [4)]
	Finally, we conduct extensive simulations to validate the advantages of the proposed energy efficiency maximization scheme for MA-aided communication systems. Simulation results show that the proposed scheme outperforms the existing MA schemes that do not account for movement-related costs and only optimize MA positions for achievable throughput or signal-to-interference-plus-noise ratio (SINR) maximization, as well as the conventional FPA scheme, in terms of users' energy efficiencies. The results also demonstrate the high robustness of the proposed scheme to imperfect angle information of multiple channel paths, thereby offering a viable solution for the practical deployment of MAs.
\end{itemize}

The rest of this paper is organized as follows. Section \ref{2} derives the energy consumption model of the stepper motor and formulates the minimum energy efficiency maximization problem. In Section \ref{3}, we first study a simple single-user scenario, which is then extended to the general multi-user case in Section \ref{4}. Next, simulation results and discussions are provided in Section \ref{5}. Finally, this paper is concluded in Section \ref{6}.

\textit{Notation:}  $a/A$, $\mathbf{a}$, $\mathbf{A}$, and $\mathcal{A}$ denote a scalar, a vector, a matrix, and a set, respectively. ${\left(  \cdot  \right)^*}$, ${\left(  \cdot  \right)^{T}}$, ${\left(  \cdot  \right)^{H}}$, $\left\|  \cdot  \right\|_2$, $\left|  \cdot  \right|$, $\mathrm{Tr}\left( \cdot\right) $, and $\mathbb{E}\left(\cdot \right) $ denote the conjugate, transpose, conjugate transpose, Euclidean norm, absolute value, trace, and expectation, respectively. $\angle x$ denotes the phase of complex number $x$. $\mathrm{diag}\left(x_1,\ldots,x_N \right) $ represents a diagonal matrix whose diagonal elements are $\left\{x_1,\ldots,x_N\right\}$. $\mathbf{I}_N$ denotes an identical matrix of size $N \times N$. $\mathbb{Z}$ denotes the set of integers. $\mathbb{C}^{M \times N}$ is the set for complex matrices of $M \times N$ dimensions. $\mathcal{CN}\left( \mathbf{0},\mathbf{\Lambda} \right) $ represents the circularly symmetric complex Gaussian (CSCG) distribution with mean zero and covariance matrix $\mathbf{\Lambda}$. $\sim$ and $\triangleq$ stand for ``distributed as'' and ``defined as'', respectively.
\section{System Model and Problem Formulation}\label{2}
\subsection{System Model}\label{system_model}
\begin{figure}[!t]
	\centering
	\includegraphics[width=1\linewidth]{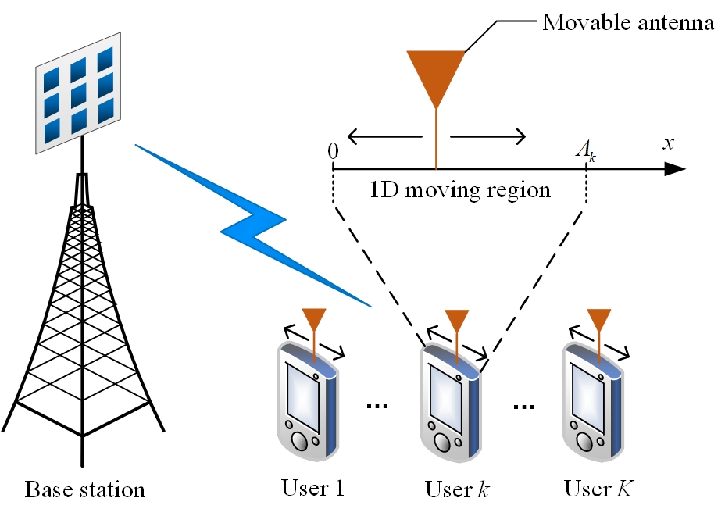}
	\caption{Illustration of the considered MA-aided uplink communication system.}
	\label{sysmodel}
\end{figure}
As shown in Fig. \ref{sysmodel}, we consider a multi-user uplink communication system, where the BS is equipped with $N$ FPAs to serve $K$ single-MA users. For each user $k$ ($k \in \mathcal{K} \triangleq \left\{1,\ldots,K \right\}$), the MA is connected to the radio frequency chain through a flexible cable with fixed length such that it can move along a linear region \cite{MA1}. Compared to conventional FPA systems, the cable length is increased only by a few wavelengths to support free MA movement, which introduces minimal feed-line loss. Furthermore, since the cable length remains constant and only one MA is connected, the variation in MA position does not significantly affect impedance matching. Hence, we assume that the MA movement has no impact on the transmit power of the users. A 1D local coordinate system in Fig. \ref{sysmodel} is established to describe the MA position for user $k$, which is denoted as $x_k \in \mathcal{A}_k$, where $\mathcal{A}_k=\left[0,A_k \right]$ is the given 1D moving region. 

According to the field-response channel model \cite{MA2}, the channel vectors of users are determined by both the propagation environment and their MA positions. Let $L_k$ denote the number of channel paths from user $k$ to the BS. Define ${\vartheta _{k,l}} = \sin {\theta _{k,l}}\cos {\phi _{k,l}}$ ($1 \le l \le L_k$) as the virtual angle of departure (AoD) of the $l$-th transmit path from user $k$, where $\theta _{k,l}$ and $\phi _{k,l}$ denote the elevation and azimuth angles, respectively. The channel vector between the BS and user $k$ can be expressed as 
\begin{equation} \label{hk}
	{\mathbf{h}_k} = \mathbf{G}_k^H{\mathbf{f}_k}\left( {{x_k}} \right),
\end{equation}
where $\mathbf{G}_k \in \mathbb{C}^{L_k \times N}$ is the path-response matrix representing the propagation environment. The entry in the $l$-th row and $n$-th ($1 \le n \le N$) column of $\mathbf{G}_k$ represents the response coefficient of the $l$-th channel path between the reference point of $\mathcal{A}_k$ and the $n$-th FPA element at the BS. In addition, ${\mathbf{f}_k}\left( {{x_k}} \right) \in \mathbb{C}^{L_k \times 1}$ is the field-response vector representing the phase differences of $L_k$ channel paths at user $k$ as a function of its MA position, which is given by
\begin{equation}
	{\mathbf{f}_k}\left( {{x_k}} \right) = {\left[ {{e^{\mathrm{j}\frac{{2\pi}}{\lambda }{x_k}{\vartheta _{k,1}}}}, \ldots ,{e^{\mathrm{j}\frac{{2\pi}}{\lambda }{x_k}{\vartheta _{k,{L_k}}}}}} \right]^T},
\end{equation}
where $\lambda$ is the carrier wavelength.

Let $\mathbf{w}_k \in \mathbb{C}^{N \times 1}$ denote the receive combining vector for user $k$. The received signal at the BS can be expressed as
\begin{equation}
	\mathbf{y} = {\mathbf{W}^H}\mathbf{H}{\mathbf{P}^{1/2}}\mathbf{s} + {\mathbf{W}^H}\mathbf{n},
\end{equation}
where $\mathbf{W}=\left[\mathbf{w}_1,\ldots,\mathbf{w}_K \right]\in \mathbb{C}^{N\times K} $ is the receive combining matrix of the BS, $\mathbf{H}=\left[\mathbf{h}_1,\ldots,\mathbf{h}_K \right]\in \mathbb{C}^{N\times K}$ denotes the channel matrix from all $K$ users to the antenna array of the BS, $\mathbf{s}\in \mathbb{C}^{K\times 1}$ represents the transmitted signals of the users, which are independent random variables with zero mean and normalized variance/power, i.e., $\mathbb{E}\left(\mathbf{ss}^H\right)=\mathbf{I}_K$, $\mathbf{P}^{1/2}=\mathrm{diag}\left(\sqrt{p_1},\ldots,\sqrt{p_K}\right)\in \mathbb{C}^{K\times K} $ is the power scaling matrix with the transmit power of user $k$ given by $p_k$, and $\mathbf{n}\in \mathbb{C}^{N\times 1} \sim \mathcal{CN}\left(\mathbf{0},\sigma^2\mathbf{I}_N\right)$ denotes the additive white Gaussian noise (AWGN) vector at the BS with covariance matrix $\sigma^2\mathbf{I}_N$. Then, the receive SINR for user $k$ at the BS can be written as
\begin{equation}\label{sinr}
	{\Gamma _k} = \frac{{{{\left| {\mathbf{w}_k^H{\mathbf{h}_k}} \right|}^2}{p_k}}}{{\sum\limits_{j = 1,j \ne k}^K {{{\left| {\mathbf{w}_k^H{\mathbf{h}_j}} \right|}^2}{p_j}}  + \left\| {{\mathbf{w}_k}} \right\|_2^2{\sigma ^2}}}. 
\end{equation}
\subsection{MA Energy Consumption Model and Energy Efficiency}
The total energy consumption of users consists of two components. The first one is the movement-related energy, which arises from the stepper motor driving the MA element to change its position. The other component is the communication-related energy, which is mainly due to the data transmission. For the movement-related energy, the stepper motor's energy consumption can be modeled by the following proposition.
\newtheorem{theorem}{Proposition}
\begin{theorem}\label{propo1}
	The stepper motor's energy consumption for the 1D movement of user $k$'s MA can be modeled as
	\begin{equation}
		E_{\mathrm{m},k} = \bar E_k\left|x_k-x^0_k \right| ,
	\end{equation}
	where $x^0_k \in \mathcal{A}_k$ is the initial position of user $k$'s MA and $\bar E_k$ represents the energy consumption rate of the user $k$'s MA movement in Joule per meter (J/m).
\end{theorem}
\begin{IEEEproof}
	We adopt a typical commercial stepper motor with two phases, i.e., phase A and phase B, and $\rho$ teeth, also known as pole pairs. As currents are applied in the proper sequence to the motor phases, the north (N) and south (S) poles of the rotor align with one or the other phase\footnote{For stepper motors with multiple phases, a dedicated current is required for each phase to ensure the continuous rotation of the rotor.}, thus producing the rotation \cite{motor1}. The electromagnetic torque of the motor can be expressed as \cite[eq. (2.6)]{motor1}
	\begin{equation}\label{torque1}
		\tau_{\mathrm {em}} = K_{\mathrm m} \left( - i_{\mathrm A}\left( t\right)  \sin \left( \rho\theta_{\mathrm m}\right)  + i_{\mathrm B}\left( t\right) \cos \left( \rho\theta_{\mathrm m} \right) \right),
	\end{equation}
	where $K_{\mathrm m}$ is the motor constant. $i_{\mathrm A}\left( t\right)$ and $i_{\mathrm B}\left( t\right)$ represent the driving currents in phase A and phase B, respectively, which vary with time $t$. $\theta_{\mathrm m}=\omega_\mathrm{m}t$ denotes the motor mechanical angle, where $\omega_\mathrm{m}$ is the rotor angular speed. Define $\omega_\mathrm{e}$ as the electrical angular speed. For continuous rotation, the duration of $\rho$ electrical periods needs to be equal to that of a single mechanical period, i.e., 
	\begin{equation}
		\rho\frac{{2\pi}}{{{\omega _\mathrm{e}}}} = \frac{{2\pi}}{{{\omega _\mathrm{m}}}} \Leftrightarrow {\omega _\mathrm{e}} = \rho{\omega _\mathrm{m}}.
	\end{equation}
	Therefore, \eqref{torque1} can be rewritten as
	\begin{equation}\label{torque2}
		\tau_{\mathrm {em}} = K_{\mathrm m} \left( - i_{\mathrm A}\left( t\right)  \sin \left( \omega _\mathrm{e}t\right)  + i_{\mathrm B}\left( t\right) \cos \left( \omega _\mathrm{e}t\right)  \right).
	\end{equation}
	To achieve a constant torque for smooth movement and high accuracy, the driving currents are typically set as
	\begin{align}
		i_{\mathrm A}\left(t \right)  &= I \cos\left(\omega_{\mathrm e} t + \frac{\pi}{2} \right), \label{ia} \\
		i_{\mathrm B}\left(t \right) & = I \cos\left(\omega_{\mathrm e} t \right),\label{ib}
	\end{align} 
	where $I$ is the peak current. By substituting \eqref{ia} and \eqref{ib} into \eqref{torque2}, we have
	\begin{equation}
		\tau_{\mathrm {em}} = K_{\mathrm m} I.
	\end{equation}
	The output energy of the stepper motor over a duration of $\hat T$ is calculated as
	\begin{equation}
		E_\mathrm{m} = \int_{0}^{\hat{T}} { \tau_{\mathrm {em}} \omega_{\mathrm m} {\mathrm d} t} =  K_{\mathrm m} I \omega_{\mathrm m} \hat{T}.
	\end{equation}
	For the proposed MA system, the duration of MA movement for user $k$ is given by ${\left| x_k-x_k^0\right| }/{v_k}$, where $v_k$ is the moving speed of user $k$'s MA, which can be interpreted as the average MA moving speed from $x_k^0$ to $x_k$ when accounting for the acceleration and deceleration effects caused by the start-stop behavior of stepper motors. Moreover, the energy conversion efficiency of user $k$'s stepper motor, denoted by $\eta_{\mathrm{m},k}$, is introduced to capture the energy consumed by the control circuit for antenna positioning and the non-linear energy losses associated with motor dynamics. Hence, the energy consumption for the user $k$'s MA movement can be derived as
	\begin{equation}\label{Emk}
		{E_{\mathrm{m},k}} = \frac{{{K_{\mathrm{m},k}}{I_k}{\omega _{\mathrm{m},k}}}}{{{\eta _{\mathrm{m},k}}}}\times \frac{{\left| {{x_k} - x_k^0} \right|}}{{{v_k}}},
	\end{equation}
	where $K_{\mathrm{m},k}$, $I_k$, and $\omega_k$ are the motor constant, peak current, and rotor angular speed of user $k$'s stepper motor, respectively. Let $r_k$ denote the rotational radius of user $k$'s stepper motor. Then, we have ${\omega _{\mathrm{m},k}}=v_k/r_k$. Therefore, \eqref{Emk} can be rewritten as
	\begin{equation}\label{Emk1}
		{E_{\mathrm{m},k}} = \frac{{{K_{\mathrm{m},k}}{I_k}{v_k}}}{{{\eta _{\mathrm{m},k}}{r_k}}} \times \frac{{\left| {{x_k} - x_k^0} \right|}}{{{v_k}}},
	\end{equation}
	where term $\left( {{{K_{\mathrm{m},k}}{I_k}{v_k}}}\right) /\left( {{{\eta _{\mathrm{m},k}}{r_k}}}\right) $ represents the power consumption of user $k$'s stepper motor, which increases linearly with ${v_k}$. By defining $\bar E_k$ as the energy consumption rate for user $k$'s stepper motor\footnote{We can see in \eqref{Emk1} that as ${v_k}$ increases, the stepper motor's power consumption $\left( {{{K_{\mathrm{m},k}}{I_k}{v_k}}}\right) /\left( {{{\eta _{\mathrm{m},k}}{r_k}}}\right) $ grows, while the MA moving delay ${\left| x_k-x_k^0\right| }/{v_k}$ decreases. Thus, we use the energy consumption rate to represent the energy required for the stepper motor to move the MA over a unit distance.}, i.e., 
	\begin{equation}\label{barE_k}
		\bar E_k\triangleq\frac{K_{\mathrm{m},k} I_k }{\eta_{\mathrm{m},k} r_k},
	\end{equation}
	\eqref{Emk1} is simplified as
	\begin{equation}
		E_{\mathrm{m},k} = \bar E_k\left|x_k-x^0_k \right|.
	\end{equation}
	This thus completes the proof.
\end{IEEEproof}
\newtheorem{remark}{Remark}
\begin{remark}
	The proposed energy consumption model for the stepper motor can be readily extended to two-dimensional (2D) or three-dimensional (3D) MA movement scenarios by employing two or three dedicated stepper motors to independently control the MA movement along orthogonal axes. Accordingly, for general 3D MA movement, the energy consumption model in Proposition \ref{propo1} can be reformulated as
	\begin{equation}
	E_{\mathrm{m},k} = \bar {E}_{\mathrm{x},k}\left|x_k-x^0_k \right|+\bar {E}_{\mathrm{y},k}\left|y_k-y^0_k \right|+\bar {E}_{\mathrm{z},k}\left|z_k-z^0_k \right|,
	\end{equation}
	where $\bar {E}_{\mathrm{x},k}$, $\bar {E}_{\mathrm{y},k}$, and $\bar {E}_{\mathrm{z},k}$ denote the energy consumption rates of MA movement along the $x$, $y$, and $z$ axes, respectively, and $x^0_k/x_k$, $y^0_k/y_k$, and $z^0_k/z_k$ represent the initial/optimized MA positions along the corresponding axes.
\end{remark}
For the communication-related energy, the energy consumption of user $k$ is given by
\begin{equation}
	{E_{\mathrm{c},k}} = T_{\mathrm{c},k}\frac{p_k}{\eta_{\mathrm{c},k}},
\end{equation}
where $T_{\mathrm{c},k}$ is the duration of communication between user $k$ and the BS, and $\eta_{\mathrm{c},k}$ denotes the user $k$'s power conversion efficiency for communication\footnote{$0 < \eta_{\mathrm{c},k}<1$ reflects the impact of the circuit power consumption for communication and can be reasonably set according to practical hardware characteristics. Moreover, if a constant circuit energy consumption is considered, the subsequent optimization problems and proposed algorithms can be modified accordingly by adding this constant energy term.}. 

During the movement of the MA, the channel may undergo severe fluctuations due to the Doppler effect. If data transmission occurs during this period, it may lead to significant degradation in QoS. As a result, it is assumed that only the control signaling is maintained between the BS and all users, while their data transmission is suspended during the MA movement. At the beginning of each transmission block, the channel state information (CSI) of the channel paths is first estimated for each position in $\mathcal{A}_k$, based on which the receive combining vectors, transmit power, and MA positions are optimized for all users. Subsequently, the users' MAs are moved to the optimized positions using stepper motors. The duration and energy consumption of CSI estimation depend on the adopted channel estimation scheme, which is beyond the scope of this paper. Hence, we do not consider the time and energy costs associated with CSI estimation and assume that perfect CSI is available for optimization. Consequently, a transmission block with duration $T$ can be divided into two phases, i.e., the MA movement phase and the communication phase. As such, the total energy consumption of user $k$ can be written as 
\begin{equation}\label{E_k}
	{E_k} = \underbrace {\bar E_k\left| {{x_k} - x_k^0} \right|}_{{\mathrm{ MA\ movement \ phase}}} + \underbrace {\frac{p_k}{\eta_{\mathrm{c},k}}\left( {T - \mathop {\max }\limits_{k \in \mathcal{K}} \left( {\frac{{\left| {{x_k} - x_k^0} \right|}}{{{v_k}}}} \right)} \right)}_{{\mathrm{Communication \ phase}}}.
\end{equation}
On the other hand, during the communication phase, the achievable throughput in bits per Hertz (bits/Hz) of user $k$ can be expressed as
\begin{equation}\label{R_k}
	{R_k} = \left( {T - \mathop {\max }\limits_{k \in \mathcal{K}} \left( {\frac{{\left| {{x_k} - x_k^0} \right|}}{{{v_k}}}} \right)} \right){\log _2}\left( {1 + {\Gamma _k}} \right).
\end{equation}
With \eqref{E_k} and \eqref{R_k}, the energy efficiency in bits per Hertz per Joule (bits/Hz/J) of user $k$ can thus be expressed as\footnote{This paper aims to characterize the energy efficiency when perfect CSI is known, which can be regarded as the performance upper bound when accounting for CSI estimation costs in practical scenarios.}
\begin{equation}
	EE_k=\frac{R_k}{E_k}.
\end{equation}
\subsection{Problem Formulation}
To guarantee fairness in QoS for each user, we aim to maximize the minimum energy efficiency among all users by jointly optimizing the receive combining matrix of the BS, as well as the transmit power and MA position of each user. Therefore, the corresponding optimization problem is formulated as
\begin{align}\label{max1}
	& \mathop {\mathrm{max} }\limits_{\mathbf{W},\left\{p_k,x_k\right\}_{k \in \mathcal{K}}} \quad \mathop {\min }\limits_{k \in \mathcal{K}} \quad EE_k \\
	&\mathrm{s.t.} \quad \mathrm{C1}:R_k \ge R_{\mathrm{TH}_k},\quad \forall k \in \mathcal{K},\nonumber\\
	&\hspace{2.3em} \mathrm{C2}: x_k \in \mathcal{A}_k,\quad \forall k \in \mathcal{K},\nonumber\\
	&\hspace{2.3em} \mathrm{C3}: 0 \le p_k \le P_{\max_k},\quad \forall k \in \mathcal{K},\nonumber
\end{align}
where constraint C1 ensures that the minimum communication throughput requirement, $R_{\mathrm{TH}_k}$, is met for user $k$, constraint C2 confines the moving region of user $k$'s MA, and constraint C3 indicates that the transmit power of user $k$ should be no larger than $P_{\max_k}$. Problem \eqref{max1} is challenging to solve because $\mathbf{W}$ and $\left\{p_k,x_k\right\}_{k \in \mathcal{K}}$ are intricately coupled in the objective function and constraint C1, which makes the problem non-convex. Thus, we first analyze the special case of the single-user scenario. Then, an efficient algorithm is proposed to solve problem \eqref{max1} for the general multi-user case.
\section{Single-User System}\label{3}
This section considers the single-user setup, i.e., $K=1$. For brevity, the user index $k$ is omitted. In this case, there is no multi-user interference, and the differences in MA moving delays for multiple users are eliminated. Since the maximum ratio combining (MRC) method is optimal for the single-user case, we have 
\begin{equation}
	\mathbf{w} = \mathbf{h}.
\end{equation}
Therefore, the user's energy efficiency is simplified as
\begin{equation}\label{EEsingle}
	EE = \frac{{\left( {T - \frac{{\left| {x - {x^0}} \right|}}{v}} \right){{\log }_2}\left( {1 + \frac{{p\left\| \mathbf{h} \right\|_2^2}}{{{\sigma ^2}}}} \right)}}{{\bar E \left| {x - {x^0}} \right| + \frac{p}{\eta_\mathrm{c}}\left( {T - \frac{{\left| {x - {x^0}} \right|}}{v}} \right)}}.
\end{equation}
To maximize the energy efficiency for the user, it is necessary to optimize its transmit power $p$ and MA position $x$. The energy efficiency maximization problem for the single-user MA system can be formulated as
\begin{align}\label{max_single}
	& \mathop {\mathrm{max} }\limits_{p,x} \quad EE \\
	&\mathrm{s.t.} \quad \mathrm{C4}: {{\left( {T - \frac{{\left| {x - {x^0}} \right|}}{v}} \right){{\log }_2}\left( {1 + \frac{{p\left\| \mathbf{h} \right\|_2^2}}{{{\sigma ^2}}}} \right)}} \ge R_\mathrm{TH},\nonumber\\
	&\hspace{2.3em} \mathrm{C5}: x \in \mathcal{A},\nonumber\\
	&\hspace{2.3em} \mathrm{C6}: 0 \le p \le P_{\max}.\nonumber
\end{align}
Since problem \eqref{max_single} is significantly simplified compared to problem \eqref{max1}, a straightforward way to solve problem \eqref{max_single} is to conduct a 1D exhaustive search. Specifically, the 1D moving region is densely discretized into $S$ sub-regions of equal length. Thus, the channel gain at the center of each sub-region, i.e., $x^{\mathrm{sub}}_s$ for $1\le s \le S$, can represent the channel gain for the entire sub-region. By optimizing transmit power $p$ for the given $x^{\mathrm{sub}}_s$ of each sub-region, the optimal solutions to problem \eqref{max_single} can be obtained. Next, we first propose an efficient algorithm to optimize $p$ for the given $x^{\mathrm{sub}}_s$. Then, to draw more insights, we derive an upper bound on the user's energy efficiency in \eqref{EEsingle} for the given optimized $p$ and illustrate the conditions required to attain this upper bound.
\subsection{Optimization Algorithm}
\begin{algorithm}[!t]
	\caption{Proposed algorithm for solving problem \eqref{max_single}}
	\label{alg1}
	\renewcommand{\algorithmicrequire}{\textbf{Initialization:}}
	\renewcommand{\algorithmicensure}{\textbf{Output:}}
	\begin{algorithmic}[1]
		\REQUIRE Set initial points $\left\{x^0,p^0\right\}$ and error tolerance $0 < \epsilon_1 \ll 1$.
		\ENSURE The optimized transmit power $p^\star$ and MA position $x^\star$.
		\FOR{$s=1:1:S$}
		\STATE Calculate the channel gain of sub-region $s$ based on its central position $x^{\mathrm{sub}}_s$ by \eqref{hk};
		\STATE Calculate the initial objective value by \eqref{EEp} and Dinkelbach variable $\alpha^0$ by \eqref{alpha};
		\STATE Initialize iteration index $i = 0$;
		\REPEAT
		\STATE Set $i=i+1$;
		\STATE Update the transmit power by \eqref{popt} for the given $\alpha^{i-1}$ and and store intermediate solution $ p^i$;
		\STATE Update the Dinkelbach variable by \eqref{alpha} for the given $ p^i$ and and store intermediate solution $\alpha^i$;
		\UNTIL {Increase of objective value \eqref{EEp} is less than $\epsilon_1$}
		\STATE Calculate objective value \eqref{EEp} for the given $x^{\mathrm{sub}}_s$ and $p^i$, i.e., $EE\left(p^i,x^{\mathrm{sub}}_s \right) $;
		\ENDFOR
		\RETURN $\left\{ {p^\star,x^\star} \right\} = \mathop {\arg \max }\nolimits_{\left\{ {{p^i},x_s^\mathrm{sub}} \right\}} \left( {EE\left( {{p^i},x_s^\mathrm{sub}} \right)} \right)$.
	\end{algorithmic}
\end{algorithm}
In this subsection, we focus on the optimization of transmit power $p$ for the given MA position $x^{\mathrm{sub}}_s$ and propose an efficient algorithm to solve problem \eqref{max_single}. By introducing Dinkelbach variable $\alpha$, and defining constants $\tilde{T} \triangleq{T - {{\left| {x - {x^0}} \right|}}/{v}} $ and $\Xi  \triangleq \bar E\left| {x - {x^0}} \right|$, the user's energy efficiency in \eqref{EEsingle} can be rewritten as
\begin{equation}\label{EEp}
	EE\left( p\right)  = \tilde T{\log _2}\left( {1 + \frac{{p{{\left\| \mathbf{h} \right\|}^2_2}}}{{{\sigma ^2}}}} \right) - \alpha \frac{{p}}{{{\eta _\mathrm{c}}}}\tilde T - \alpha \Xi,
\end{equation}
where $\alpha$ is given and can be iteratively updated as \cite{Dinkelbach}
\begin{equation}\label{alpha}
	\alpha^i = \frac{{\tilde T{{\log }_2}\left( {1 + \frac{{{p^i}{{\left\| \mathbf{h} \right\|}^2_2}}}{{{\sigma ^2}}}} \right)}}{{\Xi + \frac{{{p^i}}}{{{\eta _\mathrm{c}}}}\tilde T}}.
\end{equation}
Here, $i$ is the iteration index and $p^i$ is the optimized transmit power in the $i$-th iteration. To find the maximum value of $EE\left( p\right)$ in \eqref{EEp}, its first-order derivative is derived as 
\begin{equation}
	\dot {EE}\left( p \right) = \frac{{\tilde T{{\left\| \mathbf{h} \right\|}^2_2}}}{{\left( {p{{\left\| \mathbf{h} \right\|}^2_2} + {\sigma ^2}} \right)\ln 2}} - \alpha \frac{{\tilde T}}{{{\eta _\mathrm{c}}}}. 
\end{equation}
Let $\dot {EE}\left( p \right)=0$, we can obtain the optimal transmit power without considering constraints C4 and C6 as 
\begin{equation}
	p^\circ = \max \left(0, {\frac{{{\eta _\mathrm{c}}}}{{\alpha \ln 2}} - \frac{{{\sigma ^2}}}{{\left\| \mathbf{h} \right\|_2^2}}} \right). 
\end{equation}
To ensure that constraint C4 holds, $p$ needs to satisfy
\begin{equation}
	p \ge \frac{{{\sigma ^2}}}{{\left\| \mathbf{h} \right\|_2^2}}\left( {{2^{R_\mathrm{TH}/\tilde T}} - 1} \right) \triangleq {P_\mathrm{TH}}.
\end{equation}
As such, the optimal transmit power that maximizes the user's energy efficiency in \eqref{EEp} while satisfying constraints C4 and C6 in the $i$-th iteration is given by
\begin{equation}\label{popt}
	{p^i} = \left\{
	\begin{array}{rcl}
		P_\mathrm{TH}, & \mathrm{if} & {0\le p^\circ <P_\mathrm{TH}} ,\\
		{p^\circ}, & \mathrm{if} & P_\mathrm{TH} \le p^\circ \le P_\mathrm{max} , \\
		P_\mathrm{max}, & \mathrm{if} &p^\circ > P_\mathrm{max}  .\\
	\end{array} \right.
\end{equation}

With \eqref{alpha} and \eqref{popt}, we can alternately update $\alpha$ and $p$ until convergence is reached. The proposed algorithm for solving problem \eqref{max_single} is summarized in Algorithm \ref{alg1}.
\subsection{Performance Analysis}
Next, we analyze the upper bound on the user's energy efficiency in \eqref{EEsingle} for the given transmit power $p^\star$ optimized by Algorithm \ref{alg1}. Specifically, the upper bound on \eqref{EEsingle} can be obtained by the following proposition.
\begin{theorem} \label{proposition2}
	Define $\bar x \triangleq \mathop {\arg \max }\nolimits_{x\in \mathcal{A}} \left( {\left\| {\mathbf{h}\left( x \right)} \right\|_2^2} \right)$. If the initial MA position satisfies $x^0=\bar x$, the upper bound on \eqref{EEsingle} is achieved as
	\begin{equation}\label{ub}
		E{E^\mathrm{ub}} = \frac{\eta_\mathrm{c}}{p^\star}{{{\log }_2}\left( {1 + \frac{{p^\star\left\| {\mathbf{h}\left( {\bar x} \right)} \right\|_2^2}}{{{\sigma ^2}}}} \right)}.
	\end{equation}
\end{theorem}
\begin{IEEEproof}
	Please refer to Appendix \ref{appendixB}.
\end{IEEEproof}

Next, we conduct a detailed analysis of the conditions to achieve this performance bound under different numbers of channel paths, i.e., one-path case, two-path case, and multiple-path case. For the convenience of subsequent analysis, by denoting the entry in the $l$-th row and $n$-th column of $\mathbf{G}$ as ${{g_{ln}}}$, the closed-form expression of $\left\| \mathbf{h} \right\|_2^2$ is given by
\begin{align} \label{h22}
	\left\| \mathbf{h} \right\|_2^2 & = {\mathbf{f}^H}\left( x \right)\mathbf{G}{\mathbf{G}^H}\mathbf{f}\left( x \right) \nonumber\\
	&= X + \sum\limits_{a = 1}^{L - 1} {\sum\limits_{b = a + 1}^L {2{{\left| {{Y_{ab}}} \right|}}\cos \left( {\frac{{2\pi}}{\lambda }x{\vartheta _{ab}} + \angle  {{Y_{ab}}} } \right)} } ,
\end{align}
where $X \triangleq \sum\nolimits_{l = 1}^L {\sum\nolimits_{n = 1}^N {{{\left| {{g_{ln}}} \right|}^2}} }$, ${Y_{ab}} \triangleq \sum\nolimits_{n = 1}^N {{g_{an}}g_{bn}^*} $, and ${\vartheta _{ab}} \triangleq {\vartheta _b} - {\vartheta _a}$.
\subsubsection{One-Path Case}
For the one-path case, i.e., $L=1$, the energy efficiency in \eqref{EEsingle} can be reformulated as
\begin{equation}\label{EEL1}
	E{E_{L = 1}} = \frac{{\left( {T - \frac{{\left| {x - {x^0}} \right|}}{v}} \right){{\log }_2}\left( {1 + \frac{{p^{\star}X}}{{{\sigma ^2}}}} \right)}}{{\bar E \left| {x - {x^0}} \right| + \frac{p^{\star}}{\eta_\mathrm{c}}\left( {T - \frac{{\left| {x - {x^0}} \right|}}{v}} \right)}},
\end{equation}
which indicates that moving the MA does not improve the channel gain; instead, it leads to a reduction in communication time and an increase in the energy consumed by MA movement, thereby reducing energy efficiency. Therefore, in the one-path scenario, the optimal position of the MA is exactly its initial position.
\subsubsection{Two-Path Case}
For the two-path case, i.e., $L=2$, the channel gain in \eqref{h22} can be expressed as
\begin{equation}
	G_{L=2}\left( x \right) =X + 2\left| {{Y_{12}}} \right|\cos \left( {\frac{{2\pi}}{\lambda }x{\vartheta _{12}} + \angle {Y_{12}}} \right),
\end{equation}
which is a periodic function with period ${\lambda }/{{\left| {{\vartheta _{12}}} \right|}}$. When $x = {\lambda }\left( {d - {{\angle {Y_{12}}}}/{{2\pi}}} \right)/{{{\vartheta _{12}}}}$, $\forall d \in \mathbb{Z}$, $G_{L=2}\left( x \right)$ achieves its maximum value $G^{\max}_{L=2}=X + 2\left| {{Y_{12}}} \right|$. However, the size of the moving region, $A$, is limited. When $A \ge {\lambda }/{{\left| {{\vartheta _{12}}} \right|}}$, the upper bound on the energy efficiency is always achieved as
\begin{equation}\label{ubL2}
	EE_{L=2}^{\mathrm{ub}} = \frac{\eta_\mathrm{c}}{p^{\star}}{{{\log }_2}\left( {1 + \frac{{p^{\star}G^{\max}_{L=2}}}{{{\sigma ^2}}}} \right)},
\end{equation}
if the initial MA position satisfies $x^0 = {\lambda }\left( {d - {{\angle {Y_{12}}}}/{{2\pi}}} \right)/{{{\vartheta _{12}}}}$. When $A < {\lambda }/{{\left| {{\vartheta _{12}}} \right|}}$, the upper bound on the energy efficiency can be attained as \eqref{ubL2} if $\exists d \in \mathbb{Z}$ such that the initial MA position satisfies $0 \le x^0 = {\lambda }\left( {d - {{\angle {Y_{12}}}}/{{2\pi}}} \right)/{{{\vartheta _{12}}}} \le A$; otherwise, \eqref{ubL2} cannot be reached and the upper bound in this case is given by
\begin{equation}
	\bar{EE}_{L=2}^{\mathrm{ub}} = \frac{\eta_\mathrm{c}}{{p}^{\star}}{{{\log }_2}\left( {1 + \frac{{{p}^{\star}{\max _{x \in \mathcal{A}}}\left( {\left\| {\mathbf{h}\left( x \right)} \right\|_2^2} \right)}}{{{\sigma ^2}}}} \right)},
\end{equation}
if the initial MA position satisfies $x^0={\arg \max _{x \in \mathcal{A}}}\left( {\left\| {\mathbf{h}\left( x \right)} \right\|_2^2} \right)$.
\subsubsection{Multiple-Path Case}\label{multipath}
For the multiple-path case, i.e., $L>2$, the channel gain in \eqref{hk} can be expressed as
\begin{equation}
	G_{L>2}\left( x \right)  = \sum\limits_{n = 1}^N {\left| {\sum\limits_{l = 1}^L {g_{ln }^*{e^{\mathrm{j}\frac{{2\pi}}{\lambda }x{\vartheta _l}}}} } \right|^2}.
\end{equation}
It is difficult to explicitly represent the period of $G_{L>2}\left( x \right)$ due to the random distribution of AoDs $\theta _l$ and $\phi _l$. Nevertheless, we can approximate the period by assuming that virtual AoDs $\vartheta_l$ are quantized. Let $\chi$ denote the period of $G_{L>2}\left( x \right)$. Then, we have
\begin{align}\label{GL3}
	& G_{L>2}\left( x \right) =G_{L>2}\left( x+\chi \right) \nonumber\\
	 \Leftrightarrow & \sum\limits_{n = 1}^N {\left| {\sum\limits_{l = 1}^L {g_{ln }^*{e^{\mathrm{j}\frac{{2\pi}}{\lambda }x{\vartheta _l}}}} } \right|^2} =\sum\limits_{n = 1}^N {\left| {\sum\limits_{l = 1}^L {g_{ln }^*{e^{\mathrm{j}\frac{{2\pi}}{\lambda }\left(x+\chi \right) {\vartheta _l}}}} } \right|^2}.
\end{align}
The sufficient conditions for \eqref{GL3} to hold can be derived as
\begin{align}
	& {\left| {\sum\limits_{l = 1}^L {g_{ln }^*{e^{\mathrm{j}\frac{{2\pi}}{\lambda }x{\vartheta _l}}}} } \right|^2} = {\left| {\sum\limits_{l = 1}^L {g_{ln }^*{e^{\mathrm{j}\frac{{2\pi}}{\lambda }\left( {x + \chi } \right){\vartheta _l}}}} } \right|^2} \nonumber\\
	\Leftrightarrow & \sum\limits_{a = 1}^L {\sum\limits_{b = 1}^L {{g_{an}}g_{bn}^*{e^{\mathrm{j}\frac{{2\pi}}{\lambda }x{\vartheta _{ab}}}}} }  = \sum\limits_{a = 1}^L {\sum\limits_{b = 1}^L {{g_{an}}g_{bn}^*{e^{\mathrm{j}\frac{{2\pi}}{\lambda }\left( {x + \chi } \right){\vartheta _{ab}}}}} } \nonumber\\
	\Leftrightarrow & \sum\limits_{a = 1}^L {\sum\limits_{b = 1,b \ne a}^L {{g_{an}}g_{bn}^*{e^{\mathrm{j}\frac{{2\pi}}{\lambda }x{\vartheta _{ab}}}}\left( {1 - {e^{\mathrm{j}\frac{{2\pi}}{\lambda }\chi {\vartheta _{ab}}}}} \right)} }  = 0 \nonumber\\
	\Leftrightarrow & {\left( {1 - {e^{\mathrm{j}\frac{{2\pi}}{\lambda }\chi {\vartheta _{ab}}}}} \right)}=0, \quad 1 \le a, b \le L \nonumber\\
	\Leftrightarrow & \frac{\chi \vartheta _{ab}}{\lambda} \in \mathbb{Z}, \quad 1 \le a, b \le L,
\end{align}
which indicates that period $\chi$ is the minimum real number that ensures ${\chi \vartheta _{ab}}/{\lambda}$ to be an integer for $ 1 \le a, b \le L$. Without loss of generality, we quantize the virtual AoDs with a resolution of $Q$, i.e., ${\vartheta _l} \in {\left\{ { - 1 + {{\left( 2q - 1\right) }}/{Q}} \right\}_{1 \le q \le Q}}$, and assume that the virtual AoDs are sorted in a non-decreasing order, i.e., ${\vartheta _1} \le {\vartheta _2} \le  \ldots  \le {\vartheta _L}$. In addition, we define $\vartheta _l\triangleq { - 1 + {{\left( 2q_l - 1\right) }}/{Q}}$, which indicates that $\vartheta _l$ corresponds to the $q_l$-th element in the quantized set of virtual AoDs. Then, the difference of two adjacent virtual AoDs can be obtained as ${\vartheta _{l + 1}} - {\vartheta _l} = 2\left( {{q_{l + 1}} - {q_l}} \right)/Q \triangleq {{2{\Delta _{{q_l}}}}}/{Q}$, $1\le l \le L-1$. To guarantee ${\chi \vartheta _{ab}}/{\lambda} \in \mathbb{Z}$, the minimum period of $ G_{L>2}\left( x \right)$ should be given by 
\begin{equation}\label{period}
	\chi = \frac{Q\lambda}{2c},
\end{equation}
where $c$ is the maximal common factor for $\left\{{\Delta _{{q_l}}}\right\}_{1\le l \le L-1}$. It can be observed from \eqref{period} that the maximum value of $ G_{L>2}\left( x \right)$, i.e., $G^{\max}_{L>2}$, can always be achieved if the size of the moving region is no less than period $\chi$. As a result, if the initial MA position can be set at the location where channel gain $ G_{L>2}\left( x \right)=G^{\max}_{L>2}$, the upper bound on energy efficiency can be achieved as
\begin{equation}
	EE_{L>2}^{\mathrm{ub}} = \frac{\eta_\mathrm{c}}{p^{\star}}{{{\log }_2}\left( {1 + \frac{{p^{\star}G^{\max}_{L>2}}}{{{\sigma ^2}}}} \right)}.
\end{equation}
\section{Multi-User System}\label{4}
This section considers the general multi-user scenario\footnote{In this case, the upper bound on the minimum energy efficiency among all users can be derived based on an assumption similar to that in Proposition~\ref{proposition2}, where the initial MA position of each user is set to the location that maximizes the SINR in \eqref{sinr}.}, i.e., $K>1$. The user $k$'s energy efficiency under a given Dinkelbach variable $\alpha_k$ can be rewritten as
\begin{align}
	&EE_k = \left( {T - \mathop {\max }\limits_{k \in \mathcal{K}} \left( {\frac{{\left| {{x_k} - x_k^0} \right|}}{{{v_k}}}} \right)} \right){\log _2}\left( {1 + {\Gamma _k}} \right) \nonumber\\
	&- {\alpha _k}{{\bar E}_k}\left| {{x_k} - x_k^0} \right| - {\alpha _k}\frac{p_k}{{{\eta _{\mathrm{c},k}}}}\left( {T - \mathop {\max }\limits_{k \in \mathcal{K}} \left( {\frac{{\left| {{x_k} - x_k^0} \right|}}{{{v_k}}}} \right)} \right),
\end{align}
where $\alpha_k$ can be iteratively updated as \cite{Dinkelbach}
\begin{equation}\label{alphak}
	\alpha _k^i = \frac{{\left( {T - \mathop {\max }\limits_{k \in \mathcal{K}} \left( {\frac{{\left| {x_k^i - x_k^0} \right|}}{{{v_k}}}} \right)} \right){{\log }_2}\left( {1 + {\Gamma _k}} \right)}}{{{{\bar E}_k}\left| {x_k^i - x_k^0} \right| + \frac{p_k^i}{{{\eta _{\mathrm{c},k}}}}\left( {T - \mathop {\max }\limits_{k \in \mathcal{K}} \left( {\frac{{\left| {x_k^i - x_k^0} \right|}}{{{v_k}}}} \right)} \right)}}.
\end{equation}
Here, $p_k^i$ and $x_k^i$ represent the optimized transmit power and MA position for user $k$ in the $i$-th iteration, respectively. By introducing auxiliary variable $\varsigma$, problem \eqref{max1} can be reformulated as\footnote{This paper assumes equal user priority, which can be extended to a weighted fairness formulation by assigning user-specific thresholds $\left\{\varsigma_k\right\}_{k \in \mathcal{K}}$.}
\begin{align}\label{max2}
	& \mathop {\mathrm{max} }\limits_{\mathbf{W},\left\{p_k,x_k\right\}_{k \in \mathcal{K}},\varsigma} \quad \varsigma \\
	&\mathrm{s.t.} \quad \mathrm{C1},\mathrm{C2},\mathrm{C3}, \nonumber\\
	&\hspace{2.3em} \mathrm{C7}: EE_k \ge \varsigma,\quad \forall k \in \mathcal{K}.\nonumber
\end{align}
Problem \eqref{max2} is challenging to solve because $\mathbf{W}$ and $\left\{p_k,x_k\right\}_{k \in \mathcal{K}}$ are intricately coupled in constraints C1 and C7, which renders the problem to be non-convex. Therefore, we propose an efficient algorithm to obtain a suboptimal solution for this problem by alternately updating receive combining matrix $\mathbf{W}$, transmit power $\left\{p_k\right\}_{k \in \mathcal{K}}$, and MA positions $\left\{x_k\right\}_{k \in \mathcal{K}}$.

\subsection{Receive Combining and Transmit Power Optimization}
\begin{algorithm}[!t]
	\caption{Proposed algorithm for solving problem \eqref{max2}}
	\label{alg2}
	\renewcommand{\algorithmicrequire}{\textbf{Initialization:}}
	\renewcommand{\algorithmicensure}{\textbf{Output:}}
	\begin{algorithmic}[1]
		\REQUIRE Set initial points $\left\{{\left\{x_k^0,\varpi _k^0\right\}_{k \in \mathcal{K}},\xi _1^0}\right\}$, iteration index $i = 0$, and error tolerance $0 \le \epsilon_2 \ll 1$.
		\ENSURE The optimized receive combining matrix $\mathbf{W}$, transmit power $\left\{p_k\right\}_{k \in \mathcal{K}}$, and MA positions $\left\{x_k\right\}_{k \in \mathcal{K}}$.
		\STATE Calculate initial Dinkelbach variable $\left\{\alpha_k^0\right\}_{k \in \mathcal{K}}$ by \eqref{alphak} and receive combining matrix $\mathbf{W}^0$ by \eqref{receive_com};
		\REPEAT
		\STATE Set $i=i+1$;
		\STATE Solve problem \eqref{max4} for the given $\left\{\left\{\alpha_k^{i-1},x_k^{i-1}, \varpi _k^{i-1}\right\}_{k \in \mathcal{K}},\mathbf{W}^{i-1}\right\}$ and store the intermediate solution $\left\{{\left\{p_k^i,\bar\mu _k^i,\bar\varpi _k^i\right\}_{k \in \mathcal{K}},\bar\varsigma^i}\right\}$;
		\STATE Update $\left\{\bar\alpha^i_k\right\}_{k \in \mathcal{K}}$ and $\bar{\mathbf{W}}^i$ by \eqref{alphak} and \eqref{receive_com}, respectively;
		\STATE Solve problem \eqref{max6} for the given $\left\{\left\{\bar\alpha_k^{i},x_k^{i-1},p_k^i,\bar\mu _k^i,\bar\varpi _k^i\right\}_{k \in \mathcal{K}},\bar{\mathbf{W}}^i,\xi _1^{i-1}\right\}$ and store the intermediate solution $\left\{\left\{x_k^i,\mu _k^i,\varpi_k^i\right\}_{k \in \mathcal{K}},\xi _1^i,\xi _2^i,\varsigma^i\right\}$;	
		\STATE Update $\left\{\alpha^i_k\right\}_{k \in \mathcal{K}}$ and $\mathbf{W}^i$ by \eqref{alphak} and \eqref{receive_com}, respectively;	
		\UNTIL {Increase of objective value $\varsigma^i$ is less than $\epsilon_2$}
		\RETURN $\mathbf{W} = \mathbf{W}^i$, $\left\{p_k\right\}_{k \in \mathcal{K}}=\left\{p_k^i\right\}_{k \in \mathcal{K}}$, and $\left\{x_k\right\}_{k \in \mathcal{K}}=\left\{x_k^i\right\}_{k \in \mathcal{K}}$.
	\end{algorithmic}
\end{algorithm}
In this subsection, we optimize the receive combining matrix and transmit power for the given MA positions $\left\{x_k\right\}_{k \in \mathcal{K}}$. Note that for any given $\left\{p_k,x_k\right\}_{k \in \mathcal{K}}$, the optimal receive combining matrix can be derived in closed form based on the minimum mean square error (MMSE) receiver \cite{MA5}, i.e., 
\begin{equation}\label{receive_com}
	\mathbf{W}=\left[\mathbf{w}_1,\ldots,\mathbf{w}_K \right] = {\left( {\mathbf{H}\mathbf{P}{\mathbf{H}^H} + {\sigma ^2}{\mathbf{I}_N}} \right)^{ - 1}}\mathbf{H},
\end{equation}
where $\mathbf{w}_k = \left( {\mathbf{H}\mathbf{P}{\mathbf{H}^H} + {\sigma ^2}{\mathbf{I}_N}} \right)^{ - 1} \mathbf{h}_k$ and $\mathbf{P}=\mathrm{diag}\left(p_1,\ldots,p_K \right)$. Next, we focus on the transmit power optimization. 

To obtain a convex transmit power optimization problem, slack variables $\left\{{\mu _k},{\varpi _k}\right\}_{k \in \mathcal{K}}$ are introduced, such that
\begin{align}
	{e^{{\mu _k}}} &= \sum\limits_{j =1}^K {{{\left| {\mathbf{w}_k^H{\mathbf{h}_j}} \right|}^2}p_j} + \left\|\mathbf{w}_k\right\|_2^2\sigma^2,\quad \forall k \in \mathcal{K}, \\
	{e^{{\varpi_k}}} &= \sum\limits_{j =1,j\neq k}^K {{{\left| {\mathbf{w}_k^H{\mathbf{h}_j}} \right|}^2}p_j} + \left\|\mathbf{w}_k\right\|_2^2\sigma^2,\quad \forall k \in \mathcal{K}.
\end{align}
Since MA positions $\left\{x_k\right\}_{k \in \mathcal{K}}$ are given, we define constants $\Xi_k\triangleq{{\bar E}_k}\left| {{x_k} - x_k^0} \right|$ and $\bar T \triangleq T - \mathop {\max }\nolimits_{k \in \mathcal{K}} \left( {{\left| {{x_k} - x_k^0} \right|}/{v_k}} \right)$. As a result, problem \eqref{max2} is recast as
\begin{align}\label{max3}
	& \mathop {\mathrm{max} }\limits_{\left\{p_k,\mu _k,\varpi_k\right\}_{k \in \mathcal{K}},\varsigma} \quad \varsigma \\
	&\mathrm{s.t.} \quad \mathrm{C3}, \nonumber\\
	&\hspace{2.3em}\bar{\mathrm{C1}}:\bar T{\log _2} {{e^{\left( {{\mu _k} - {\varpi _k}} \right)}}}  \ge {R_{\mathrm{TH}_k}},\quad \forall k \in \mathcal{K}, \nonumber\\
	&\hspace{2.3em} \bar{\mathrm{C7}}: \bar T{\log _2}{e^{\left( {{\mu _k} - {\varpi _k}} \right)}}  - {\alpha _k}\bar T\frac{{p_k}}{{{\eta _{\mathrm{c},k}}}}- {\alpha _k}{\Xi _k} \ge \varsigma, \forall k \in \mathcal{K} ,\nonumber\\
	&\hspace{2.3em} \mathrm{C8}: \sum\limits_{j =1}^K {{{\left| {\mathbf{w}_k^H{\mathbf{h}_j}} \right|}^2}p_j} + \left\|\mathbf{w}_k\right\|_2^2\sigma^2 \ge {e^{{\mu _k}}},\quad \forall k \in \mathcal{K} ,\nonumber\\
	&\hspace{2.3em} \mathrm{C9}: \sum\limits_{j =1,j\neq k}^K {{{\left| {\mathbf{w}_k^H{\mathbf{h}_j}} \right|}^2}p_j} + \left\|\mathbf{w}_k\right\|_2^2\sigma^2 \le {e^{{\varpi_k}}}, \forall k \in \mathcal{K} .\nonumber\
\end{align}
However, problem \eqref{max3} is still non-convex due to the non-convex constraint C9. By applying the first-order Taylor expansion, the lower bound surrogate function for ${e^{{\varpi_k}}}$ can be derived as ${e^{\varpi _k^i}} + {e^{\varpi _k^i}}\left( {{\varpi _k} - \varpi _k^i} \right)$, where $\left\{\varpi _k^i\right\}_{k \in \mathcal{K}}$ are the given local points in the $i$-th iteration. Thus, constraint C9 can be restated as
\begin{align}
	\bar{\mathrm{C9}}: \sum\limits_{j =1,j\neq k}^K &{{{\left| {\mathbf{w}_k^H{\mathbf{h}_j}} \right|}^2}p_j} + \left\|\mathbf{w}_k\right\|_2^2\sigma^2 \nonumber\\
	&\le {e^{\varpi _k^i}} + {e^{\varpi _k^i}}\left( {{\varpi _k} - \varpi _k^i} \right), \quad \forall k \in \mathcal{K}.
\end{align}
As a result, problem \eqref{max3} is transformed into the following convex semidefinite program problem: 
\begin{align}\label{max4}
	& \mathop {\mathrm{max} }\limits_{\left\{p_k,\mu _k,\varpi_k\right\}_{k \in \mathcal{K}},\varsigma} \quad \varsigma \\
	&\mathrm{s.t.} \quad \bar{\mathrm{C1}},\mathrm{C3},\bar{\mathrm{C7}},\mathrm{C8},\bar{\mathrm{C9}}, \nonumber
\end{align}
which can be efficiently solved using standard convex solvers such as CVX \cite{CVX}.
\subsection{Antenna Position Optimization}
In this subsection, we optimize the MA positions for the given receive combining matrix $\mathbf{W}$ and transmit power $\left\{p_k\right\}_{k \in \mathcal{K}}$. By defining slack variables $\left\{{\xi _1},{\xi _2}\right\}$, the antenna position optimization problem is formulated as
\begin{align}\label{max5}
	& \mathop {\mathrm{max} }\limits_{\left\{x_k,\mu _k,\varpi_k\right\}_{k\in \mathcal{K}},\varsigma,\xi _1,\xi _2} \quad \varsigma \\
	&\mathrm{s.t.} \quad \mathrm{C2}, \nonumber\\
	&\hspace{2.3em}\mathrm{C10}: {\xi _1} \ge \frac{{{x_k} - x_k^0}}{{{v_k}}},\quad{\xi _1} \ge \frac{{x_k^0-{x_k}}}{{{v_k}}},\quad \forall k \in \mathcal{K} ,\nonumber\\
	&\hspace{2.3em} \mathrm{C11}: {\xi _2} \le \frac{{{x_k} - x_k^0}}{{{v_k}}},\quad{\xi _2} \le \frac{{x_k^0-{x_k}}}{{{v_k}}},\quad \forall k \in \mathcal{K} ,\nonumber\\
	&\hspace{2.3em} \mathrm{C12}: \left( {T - {\xi _1}} \right){\log _2}{e^{\left( {{\mu _k} - {\varpi _k}} \right)}} \ge {R_{\mathrm{TH}_k}},\quad \forall k \in \mathcal{K} ,\nonumber\\
	&\hspace{2.3em} \mathrm{C13}: \left( {T - {\xi _1}} \right){\log _2}{e^{\left( {{\mu _k} - {\varpi _k}} \right)}} - {\alpha _k}{{\bar E}_k}{\xi _1}{v_k} \nonumber\\
	&\hspace{4.8em}- {\alpha _k}\frac{{p_k }}{{{\eta _{\mathrm{c},k}}}}\left( {T - {\xi _2}} \right) \ge \varsigma ,\quad \forall k \in \mathcal{K},\nonumber\\
	&\hspace{2.3em} \mathrm{C14}: \sum\limits_{j=1}^K {\mathbf{h}_j^H{\mathbf{w}_k\mathbf{w}_k^H}{\mathbf{h}_j}p_j}  + \left\|\mathbf{w}_k\right\|_2^2\sigma^2 \ge {e^{{\mu _k}}}, \forall k \in \mathcal{K} ,\nonumber\\
	&\hspace{2.3em} \mathrm{C15}: \sum\limits_{j=1,j \neq k}^K {\mathbf{h}_j^H{\mathbf{w}_k\mathbf{w}_k^H}{\mathbf{h}_j}p_j}  + \left\|\mathbf{w}_k\right\|_2^2\sigma^2 \nonumber\\
	&\hspace{4.8em}\le {e^{\varpi _k^i}} + {e^{\varpi _k^i}}\left( {{\varpi _k} - \varpi _k^i} \right),\quad \forall k \in \mathcal{K}.\nonumber\
\end{align}
Problem \eqref{max5} is hard to solve because variables $\left\{\left\{\mu _k,\varpi _k\right\}_{k\in\mathcal{K}},\xi _1\right\}$ are coupled in constraints C12 and C13, and constraints C14 and C15 are non-convex with respect to $\left\{x_k\right\}_{k \in \mathcal{K}}$. Therefore, we propose an SCA-based algorithm to solve this problem.

For constraints C12 and C13 with coupled variables $\left\{\left\{\mu _k,\varpi _k\right\}_{k \in \mathcal{K}},\xi _1\right\}$, we have \cite[eqs. (101) and (102)]{SCA}
\begin{align}
	{\xi _1}{\mu _k} &\le \frac{1}{2}\left( {\frac{{\mu _k^i}}{{\xi _1^i}}\xi _1^2 + \frac{{\xi _1^i}}{{\mu _k^i}}\mu _k^2} \right),\quad \forall k \in \mathcal{K},\label{40}\\
	{\xi _1}{\varpi _k} & \ge \left( {1 + \ln {\xi _1} + \ln {\varpi _k} - \ln \xi _1^i - \ln \varpi _k^i} \right)\xi _1^i\varpi _k^i, \forall k \in \mathcal{K},\label{41}
\end{align}
where $\left\{\left\{\mu _k^i,\varpi _k^i\right\}_{k\in\mathcal{K}},\xi _1^i\right\}$ are the given local points in the $i$-th iteration. Hence, the convex approximations of constraints C12 and C13 can be respectively formulated as
\begin{align}
	 \bar{\mathrm{C12}}:& T\left( {{\mu _k} - {\varpi _k}} \right) - \frac{1}{2}\left( {\frac{{\mu _k^i}}{{\xi _1^i}}\xi _1^2 + \frac{{\xi _1^i}}{{\mu _k^i}}\mu _k^2} \right) \nonumber\\
	 &+ \left( {1 + \ln {\xi _1} + \ln {\varpi _k} - \ln \xi _1^i - \ln \varpi _k^i} \right)\xi _1^i\varpi _k^i \nonumber\\
	 &\ge {R_{\mathrm{TH}_k}}\ln 2,\quad \forall k \in \mathcal{K}, \\
	 \bar{\mathrm{C13}}:&T\left( {{\mu _k} - {\varpi _k}} \right) - \frac{1}{2}\left( {\frac{{\mu _k^i}}{{\xi _1^i}}\xi _1^2 + \frac{{\xi _1^i}}{{\mu _k^i}}\mu _k^2} \right) \nonumber\\
	 &+ \left( {1 + \ln {\xi _1} + \ln {\varpi _k} - \ln \xi _1^i - \ln \varpi _k^i} \right)\xi _1^i\varpi _k^i \nonumber\\
	 &- \left( {{\alpha _k}{{\bar E}_k}{\xi _1}{v_k} + {\alpha _k}\frac{{p_k}}{{{\eta _{\mathrm{c},k}}}}\left( {T - {\xi _2}} \right)} \right)\ln 2 \nonumber\\
	 &\ge \varsigma \ln 2,\quad \forall k \in \mathcal{K}.
\end{align}

To make $\left\{x_k\right\}_{k \in \mathcal{K}}$ explicit in constraints C14 and C15, the term ${\mathbf{h}_j^H{\mathbf{w}_k}{\mathbf{w}_k^H}{\mathbf{h}_j}p_j}$ is rewritten as
\begin{align}\label{hkj}
	&{h_{jk}}\left( {{x_j}} \right)  = {\mathbf{h}_j^H{\mathbf{w}_k}{\mathbf{w}_k^H}{\mathbf{h}_j}p_j} \nonumber\\
	&= \mathbf{f}_j^H\left( {{x_j}} \right){\mathbf{G}_j}{\mathbf{w}_k}{\mathbf{w}_k^H}\mathbf{G}_j^H p_j{\mathbf{f}_j}\left( {{x_j}} \right)  =\mathrm{Tr}\left( {{\mathbf{M}_{jk}}} \right)\nonumber\\
	& + \sum\limits_{a = 1}^{{L_k} - 1} {\sum\limits_{b = a + 1}^{{L_k}} 2{\left| {{m_{jk,ab}}} \right|\cos \left( {\frac{{2\pi }}{\lambda }{x_j}{\vartheta _{k,ab}} + \angle {m_{jk,ab}}} \right)} },
\end{align}
where we define $\mathbf{M}_{jk}\triangleq{\mathbf{G}_j}{\mathbf{w}_k}{\mathbf{w}_k^H}\mathbf{G}_j^H p_j \in \mathbb{C}^{L_k \times L_k}$ and $\vartheta _{k,ab}\triangleq\vartheta _{k,b}-\vartheta _{k,a}$, and ${m_{jk,ab}}$ represents the entry in the $a$-th row and $b$-th column of ${\mathbf{M}_{jk}}$. To address the non-convex constraints C14 and C15, we apply the SCA algorithm based on \eqref{hkj}. Specifically, for the given local points $\left\{x_j^i\right\}_{j\in \mathcal{K}}$ in the $i$-th iteration, the lower bound and upper bound quadratic surrogate functions for ${h_{jk}}\left( {{x_j}} \right)$ can be respectively constructed as \cite{MA3}
\begin{align}
	&{h_{jk}}\left( {{x_j}} \right) \ge h_{jk}^{\mathrm{lb},i}\left( {{x_j}} \right) \nonumber\\
	&= {h_{jk}}\left( {x_j^i} \right) + \frac{{\mathrm{d}{h_{jk}}\left( {x_j^i} \right)}}{{\mathrm{d}{x_j}}}\left( {{x_j} - x_j^i} \right) - \frac{{{\varepsilon _{jk}}}}{2}{\left( {{x_j} - x_j^i} \right)^2} ,\label{45}\\
	&{h_{jk}}\left( {{x_j}} \right) \le h_{jk}^{\mathrm{ub},i}\left( {{x_j}} \right) \nonumber\\
	&= {h_{jk}}\left( {x_j^i} \right) + \frac{{\mathrm{d}{h_{jk}}\left( {x_j^i} \right)}}{{\mathrm{d}{x_j}}}\left( {{x_j} - x_j^i} \right) + \frac{{{\varepsilon _{jk}}}}{2}{\left( {{x_j} - x_j^i} \right)^2},\label{46}
\end{align}
where ${{\mathrm{d}{h_{jk}}\left( {x_j^i} \right)}}/{{\mathrm{d}{x_j}}}$ is given by
\begin{small}
\begin{align}\label{1order}
	&\frac{{\mathrm{d}{h_{jk}}\left( {x_j^i} \right)}}{{\mathrm{d}{x_j}}} \nonumber\\
	&= \sum\limits_{a = 1}^{{L_k}-1} {\sum\limits_{b =a+1}^{{L_k}} { - \frac{{4\pi\left| {{m_{jk,ab}}} \right|{\vartheta _{k,ab}}}}{\lambda }\sin \left( {\frac{{2\pi}}{\lambda }x_j^i{\vartheta _{k,ab}} + \angle {m_{jk,ab}}} \right)} } ,
\end{align}
\end{small}and $\varepsilon _{jk}$ is a positive real number satisfying $\varepsilon _{jk} \ge {{{\mathrm{d}^2}h_{jk}\left( x_j \right)}}/{{\mathrm{d}{x_j^2}}}$, with the closed-form expression given in \eqref{varkj} of Appendix \ref{appendixC}. Thus, constraints C14 and C15 can be respectively approximated as
\begin{align}
	&\bar{\mathrm{C14}}:\sum\limits_{j=1}^K {h_{jk}^{\mathrm{lb},i}\left( {{x_j}} \right)}  + \left\|\mathbf{w}_k\right\|_2^2\sigma^2 \ge {e^{{\mu _k}}} ,\quad \forall k \in \mathcal{K},\\
	&\bar{\mathrm{C15}}:\sum\limits_{j=1,j \neq k}^K {h_{jk}^{\mathrm{ub},i}\left( {{x_j}} \right)}  + \left\|\mathbf{w}_k\right\|_2^2\sigma^2 \nonumber\\
	&\hspace{2.5em}\le {e^{\varpi _k^i}} + {e^{\varpi _k^i}}\left( {{\varpi _k} - \varpi _k^i} \right), \quad \forall k \in \mathcal{K}.
\end{align}

After addressing the non-convex constraints C12, C13, C14, and C15, problem \eqref{max5} can be reformulated as the following convex optimization problem:
\begin{align}\label{max6}
	& \mathop {\mathrm{max} }\limits_{\left\{x_k,\mu _k,\varpi_k\right\}_{k\in \mathcal{K}},\varsigma,\xi _1,\xi _2} \quad \varsigma \\
	&\mathrm{s.t.} \quad \mathrm{C2},\mathrm{C10},\mathrm{C11},\bar{\mathrm{C12}},\bar{\mathrm{C13}},\bar{\mathrm{C14}},\bar{\mathrm{C15}}, \nonumber
\end{align}
which can be optimally solved by CVX \cite{CVX}. The proposed algorithm for solving problem \eqref{max2} is summarized in Algorithm~\ref{alg2}.
\subsection{Convergence and Complexity Analysis}
Let $\varsigma_1 $ and $\varsigma_2$ denote the objective values of problems \eqref{max4} and \eqref{max6}, respectively. Therefore, in lines 3-7 of Algorithm \ref{alg2}, we have
\begin{align}
	&{\varsigma _1}\left( {{{\left\{ {p_k^i,\bar \mu _k^i,\bar \varpi _k^i,\alpha _k^{i - 1},x_k^{i - 1}} \right\}}_{k \in \mathcal{K}}}},\mathbf{W}^{i-1} \right) \nonumber\\
	\mathop  \le \limits^{\left( {{a_1}} \right)} &{\varsigma _1}\left( {{{\left\{ {p_k^i,\bar \mu _k^i,\bar \varpi _k^i,\bar \alpha _k^i,x_k^{i - 1}} \right\}}_{k \in \mathcal{K}}}},\bar{\mathbf{W}}^{i} \right)\nonumber\\
	\mathop  = \limits^{\left( {{a_2}} \right)} &{\varsigma _2}\left( {{{\left\{ {p_k^i,\bar \mu _k^i,\bar \varpi _k^i,\bar \alpha _k^i,x_k^{i - 1}} \right\}}_{k \in \mathcal{K}}},\bar{\mathbf{W}}^{i},\xi _1^{i - 1},\xi _2^{i - 1}} \right)\nonumber\\
	\mathop  \le \limits^{\left( {{a_3}} \right)} &{\varsigma _2}\left( {{{\left\{ {p_k^i,\mu _k^i,\varpi _k^i,\bar \alpha _k^i,x_k^i} \right\}}_{k \in \mathcal{K}}},\bar{\mathbf{W}}^{i},\xi _1^i,\xi _2^i} \right)\nonumber\\
	\mathop  \le \limits^{\left( {{a_4}} \right)} &{\varsigma _2}\left( {{{\left\{ {p_k^i,\mu _k^i,\varpi _k^i,\alpha _k^i,x_k^i} \right\}}_{k \in \mathcal{K}}},\mathbf{W}^{i},\xi _1^i,\xi _2^i} \right)\nonumber\\
	\mathop  = \limits^{\left( {{a_5}} \right)} &{\varsigma _1}\left( {{{\left\{ {p_k^i,\mu _k^i,\varpi _k^i,\alpha _k^i,x_k^i} \right\}}_{k \in \mathcal{K}}}},\mathbf{W}^{i} \right)\nonumber\\
	\mathop  \le \limits^{\left( {{a_6}} \right)} &{\varsigma _1}\left( {{{\left\{ {p_k^{i + 1},\bar \mu _k^{i + 1},\bar \varpi _k^{i + 1},\alpha _k^i,x_k^i} \right\}}_{k \in \mathcal{K}}}},\mathbf{W}^{i} \right). 
\end{align}
The inequalities marked by $\left( {{a_1}} \right)$ and $\left( {{a_4}} \right)$ hold because $\left\{\bar \alpha _k^i, \alpha _k^i\right\}_{k \in \mathcal{K}}$ and $\left\{\bar{\mathbf{W}}^{i},\mathbf{W}^{i}\right\}$ are Dinkelbach variables and optimal receive combining matrices updated according to \eqref{alphak} and \eqref{receive_com}, respectively. The equality marked by $\left( {{a_2}} \right)$ holds because constraints C10 and C11 are active at given $\left\{\left\{x_k^{i-1}\right\}_{k \in \mathcal{K}},\xi_1^{i-1},\xi_2^{i-1}\right\}$, i.e., $\xi_1^{i-1}=\xi_2^{i-1}=\mathop {\max }\nolimits_{k \in \mathcal{K}} \left( {{\left| {{x_k^{i-1}} - x_k^0} \right|}/{v_k}} \right)$, and convex approximations \eqref{40} and \eqref{41} as well as second-order Taylor expansions \eqref{45} and \eqref{46} are all tight at given $\left\{ {{{\left\{ {p_k^i,\bar \mu _k^i,\bar \varpi _k^i,\bar \alpha _k^i,x_k^{i - 1}} \right\}}_{k \in \mathcal{K}}},\bar{\mathbf{W}}^{i},\xi _1^{i - 1},\xi _2^{i - 1}} \right\}$. The inequalities marked by $\left( {{a_3}} \right)$ and $\left( {{a_6}} \right)$ hold because $\left\{ {{{\left\{ {\mu _k^i,\varpi _k^i,x_k^i} \right\}}_{k \in \mathcal{K}}},\xi _1^i,\xi _2^i} \right\}$ and ${{{\left\{ {p_k^{i + 1},\bar \mu _k^{i + 1},\bar \varpi _k^{i + 1}} \right\}}_{k \in \mathcal{K}}}}$ are the optimal solutions to problems \eqref{max4} and \eqref{max6}, respectively. The equality marked by $\left( {{a_5}} \right)$ holds because problem \eqref{max4} is feasible and has the same objective value as problem \eqref{max6} at given $\left\{ {{{\left\{ {p_k^i,\mu _k^i,\varpi _k^i,\alpha _k^i,x_k^i} \right\}}_{k \in \mathcal{K}}}},\mathbf{W}^{i} \right\}$. As such, the objective value of problem \eqref{max6} is non-decreasing during the iterations in lines 3-7 of Algorithm \ref{alg2}. Besides, the optimal objective value of problem \eqref{max6} is upper-bounded due to the minimum communication throughput constraints. Consequently, the convergence of Algorithm \ref{alg2} is guaranteed.

The computational complexity of Algorithm \ref{alg2} primarily arises from solving problems \eqref{max4} and \eqref{max6}, with corresponding complexities of $\mathcal{O}\left( {{K^{3.5}}{N^{6.5}}\ln {\epsilon_2^{-1} }} \right)$ and $\mathcal{O}\left( {{K^{3.5}}\ln {\epsilon_2^{-1} }} \right)$. Thus, the computational complexity of Algorithm \ref{alg2} is approximately $\mathcal{O}\left( {\bar I{K^{3.5}}{N^{6.5}}\ln {\epsilon_2^{-1} }} \right)$, where $\bar I$ denotes the number of iterations needed to achieve convergence\footnote{To reduce the computational complexity for real-time scheduling, the number of iterations in Algorithm \ref{alg2} can be limited, with a moderate loss in system performance.}. 
\subsection{Implementation Considerations}
In this subsection, we discuss some practical considerations for the implementation of the proposed MA system.
\subsubsection{Channel Estimation}
It is essential to acquire accurate CSI between the BS and users to determine favorable MA positions. In existing channel estimation schemes for MA systems \cite{add_channel1,add_channel2}, the MA needs to be moved to pre-designed positions along different trajectories for channel measurements, which consumes additional energy and time. As a result, the considered system is more suitable for slowly varying channels with static or low-mobility terminals, where CSI does not need to be frequently refreshed, such as in machine-type communication (MTC) scenarios. Moreover, the robustness of the considered system to imperfect CSI will be demonstrated through simulations in the next section.
\subsubsection{Discrete MA Movement}
In practical MA systems, due to hardware constraints such as stepper motor resolution and mechanical control accuracy, MA movement may be restricted to a small fixed step size. In this case, the MA positions optimized by the proposed algorithms should be projected onto the nearest discrete positions for implementation. In addition, several dedicated algorithms \cite{6DMA2,add1,add2} have been developed to optimize discrete MA positions for finite-precision antenna positioning in practice.
\section{Simulation Results}\label{5}
\renewcommand{\arraystretch}{1.3}
\begin{table}[!t]
	\caption{Simulation Parameters}
	\label{tab1}
	\centering
	\begin{tabular}{|l|l|l|}
		\hline
		\multicolumn{1}{|c|}{\textbf{Parameter}} & \multicolumn{1}{c|}{\textbf{Description}} & \multicolumn{1}{c|}{\textbf{Value}} \\ \hline
		$\lambda$ & Carrier wavelength & 0.01 m \\ \hline
		$A$ & Length of moving region & $\lambda$ \\ \hline
		$N$ & Number of antennas at the BS & 16 \\ \hline
		$K$ & Number of users & 4 \\ \hline
		$L$ & Number of channel paths & 10 \\ \hline
		$\rho_0$ & Path loss at the reference distance & $-40$ dB \\ \hline
		$d$ & Distance between the BS and users & 50 m \\ \hline
		$\tau$ & Path loss exponent & 2.8 \\ \hline
		$P_{\mathrm{max}}$ & Maximum transmit power & 10 dBm   \\ \hline
		$\eta_{\mathrm{c},k}$ & Energy conversion efficiency of user $k$ & 50\%   \\ \hline
		$\bar E$ & Energy consumption rate of MA movement & 0.175 J/m   \\ \hline
		$v$ & MA moving speed & 0.1 m/s   \\ \hline
		$T$ & Transmission block duration & 2 s   \\ \hline
		$R_\mathrm{TH}$ & Minimum throughput requirement & 0.8 bits/Hz   \\ \hline
		$\sigma^2$ & Average noise power & $-70$ dBm   \\ \hline
		$\epsilon_1$, $\epsilon_2$ & Error tolerance  & $10^{-6}$   \\ \hline
	\end{tabular}
\end{table}
This section presents extensive simulation results to validate the effectiveness of the proposed schemes. Without loss of generality, the users are assumed to be uniformly distributed around the BS at a distance $d$. We assume that the number of channel paths for each user is identical, i.e., $L_k=L$, and adopt the channel model in \eqref{hk}, where each element in $\mathbf{G}_k$ follows CSCG distribution $\mathcal{CN}\left(0,\rho_0d^{-\tau}/L \right) $. Here, $\rho_0$ denotes the path loss at the reference distance of 1 meter (m) and $\tau$ represents the path loss exponent. The elevation and azimuth AoDs of the channel paths for each user are assumed to be independent and identically distributed random variables\footnote{The randomness is justified because each plane wave in \eqref{hk} may be composed of multiple unresolvable reflected paths within the scatterer \cite{Th}. To further account for the strong spatial correlation introduced by linear movement, future work may incorporate corresponding AoD distributions or leverage channel-sounder data accordingly \cite{add3}.} within the interval $\left[0,\pi\right]$. We also assume that all users employ stepper motors with identical energy consumption rates, i.e., $\bar E_k=\bar E$, and their minimum communication throughput requirements and maximum transmit power are identical, i.e., $R_{\mathrm{TH}_k}=R_{\mathrm{TH}}$ and $P_{\max_k}=P_{\max}$, respectively. Besides, the moving speeds, moving region sizes, and initial positions of the transmit MAs at users are set identically as $v_k=v$, $\mathcal{A}_k=\mathcal{A}=\left[0,A\right]$, and $x^0_k=x^0=A/2$, respectively. Unless otherwise specified, the default simulation parameters are listed in Table \ref{tab1} based on existing literature \cite{MA5,Th,energy1} and typical parameters of commercial stepper motors \cite{motor1,motor2}.
\subsection{Single-User System}\label{simu_sg}
First, we consider the single-user setup, i.e., $K=1$. The user's 1D moving region is discretized into multiple sub-regions of length $\lambda/100$ for the 1D exhaustive search. The results obtained by Algorithm \ref{alg1} are termed as \textit{Proposed}. Besides, the following five benchmark schemes are considered for performance comparison: 1) \textit{Upper bound}: The upper bound is given in Proposition \ref{proposition2}; 2) \textit{Quantized}: The AoDs at the user are quantized with a resolution of $Q$ according to Section \ref{multipath}, based on which Algorithm \ref{alg1} is used to optimize the MA position; 3) \textit{Max throughput}: The user's MA position is optimized to maximize the achievable throughput within the given transmission block duration, and the transmit power is optimized to maximize the energy efficiency; 4) \textit{Max SNR}: The user's MA position is optimized to maximize the channel gain, i.e., the receive signal-to-noise ratio (SNR) of the signal from the user, and the transmit power is optimized to maximize the energy efficiency; 5) \textit{FPA}: The user's antenna position is fixed at the initial position, and the transmit power is optimized to maximize the energy efficiency.

\begin{figure}[!t]
	\centering
	\subfloat[]{\label{A_EE}\includegraphics[width=1\columnwidth]{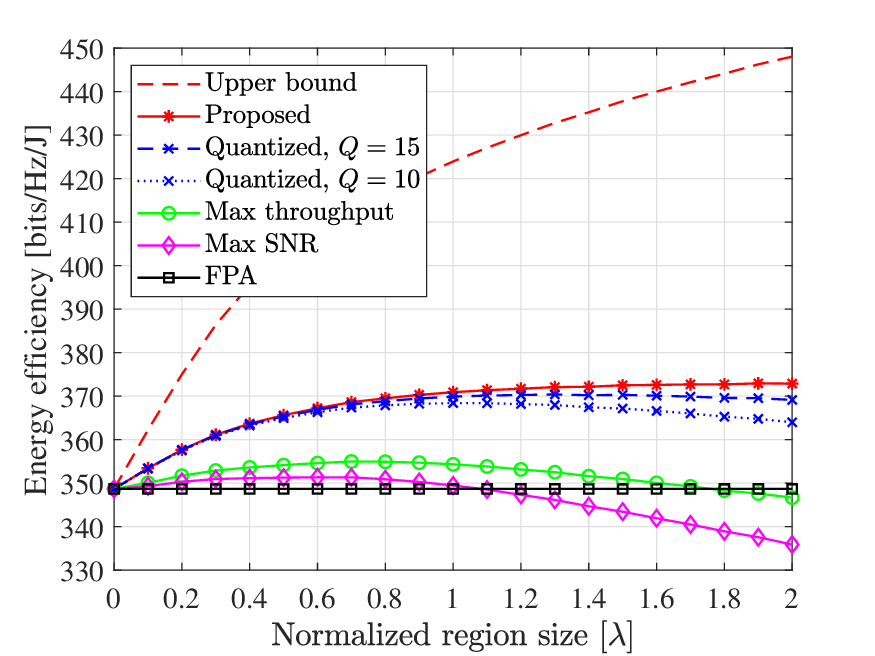}}\\
	\subfloat[]{\label{A_mov_dis}\includegraphics[width=1\columnwidth]{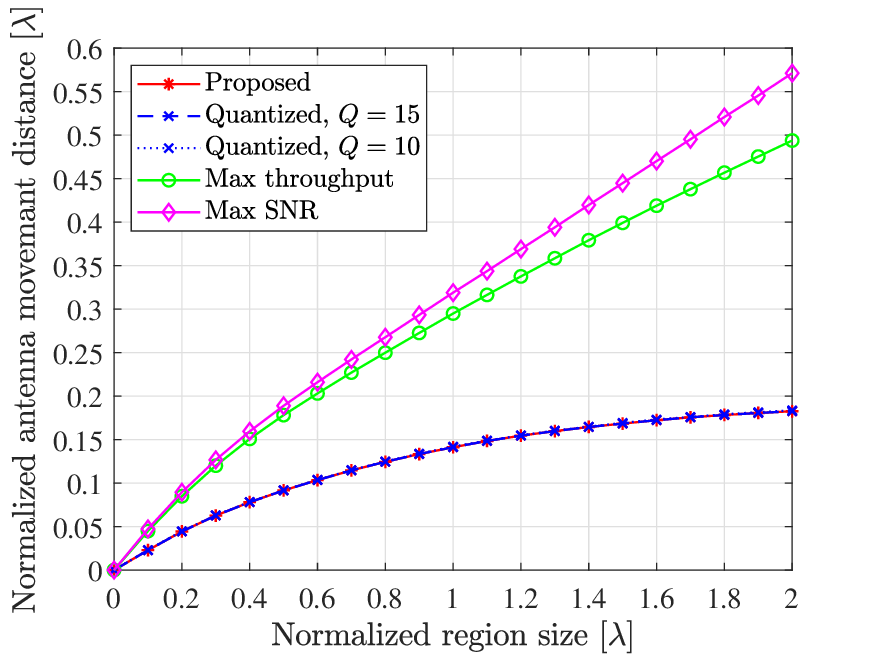}}
	\caption{Energy efficiency and normalized MA moving distance versus normalized region size.}
	\label{sg_A}
\end{figure}
Fig. \ref{sg_A} shows the energy efficiencies and normalized MA moving distances of different schemes versus the normalized region size, where the MA moving distance and the length of moving region are normalized by carrier wavelength, i.e., $\left( x^\star-x^0\right)/\lambda$ and $A/\lambda$, respectively. We can see in Fig. \ref{sg_A}\subref{A_EE} that the \textit{Upper bound} increases as the moving region expands. This is because a larger moving region allows the MA to better exploit the spatial DoFs, thereby identifying positions with higher channel gain. Besides, the upper bound in \eqref{ub} does not account for any movement-related energy or time consumption. However, the \textit{Upper bound} is overly idealized, as in practical scenarios, the user's initial MA position cannot always be at the position with the maximum channel gain. More practically, the \textit{Proposed} scheme comprehensively balances the trade-off between improving communication quality and reducing MA moving overhead. Therefore, its energy efficiency generally improves as $A/\lambda$ increases and eventually stabilizes for large $A/\lambda$. The \textit{Quantized} scheme shows a slight decrease in energy efficiency at large values of $A/\lambda$, and this decline becomes more significant as quantization resolution $Q$ decreases. This is because when a small $Q$ is used, the difference between the practical AoDs and the approximately quantized AoDs becomes large, thereby leading to the inaccurate period of channel gain, particularly in large moving regions that encompass multiple periods. Unlike the results in existing MA works, the energy efficiencies of the \textit{Max throughput} and \textit{Max SNR} schemes do not always increase with the enlargement of the moving region. This is particularly evident in the \textit{Max SNR} scheme, where a sharp decline in energy efficiency is observed when $A/\lambda > 0.8$. More critically, the energy efficiencies of the \textit{Max throughput} and \textit{Max SNR} schemes fall below that of the \textit{FPA} scheme when $A/\lambda > 1.8$ and $A/\lambda > 1.2$, respectively. This is because as the length of the MA moving region increases, the position with the maximum channel gain may be farther from the initial MA position. The long-distance MA movement leads to high energy consumption by the stepper motor and also severely squeezes the duration of the communication phase within a given transmission block. As shown in Fig. \ref{sg_A}\subref{A_mov_dis}, the \textit{Max SNR} scheme optimizes the MA position solely for maximum channel gain, which results in a linear increase in the MA moving distance as $A/\lambda$ increases. The \textit{Max throughput} scheme, which considers the delay caused by MA movement to maximize achievable throughput, has a shorter MA moving distance than the \textit{Max SNR} scheme. Thanks to the consideration of both the energy and delay overheads associated with MA movement, the \textit{Proposed} and \textit{Quantized} schemes maintain a stable MA moving distance when $A/\lambda$ is large, which contributes to superior energy efficiency.

\begin{figure}[!t]
	\centering
	\subfloat[]{\label{E_MA_EE}\includegraphics[width=1\columnwidth]{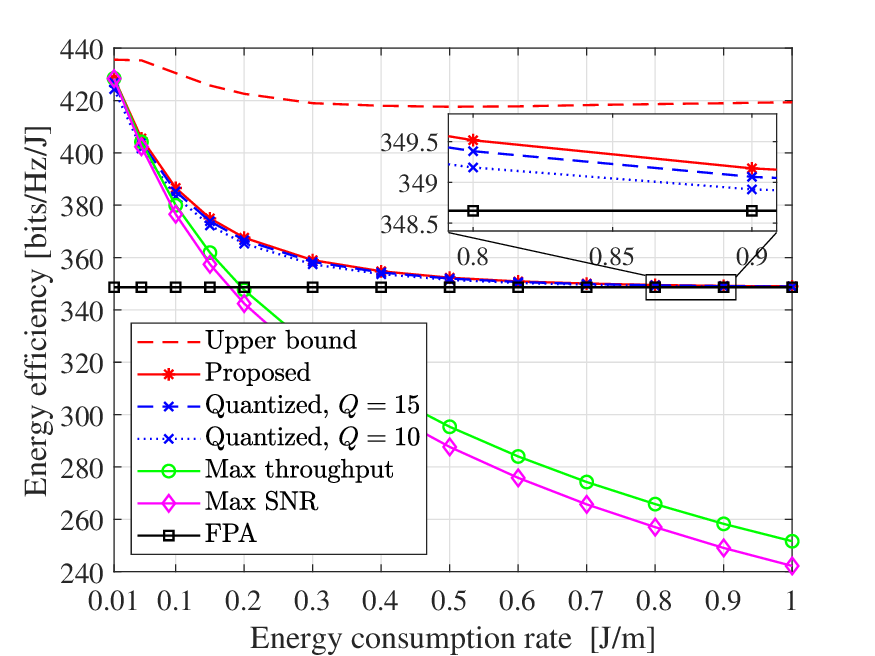}}\\
	\subfloat[]{\label{E_MA_mov_dis}\includegraphics[width=1\columnwidth]{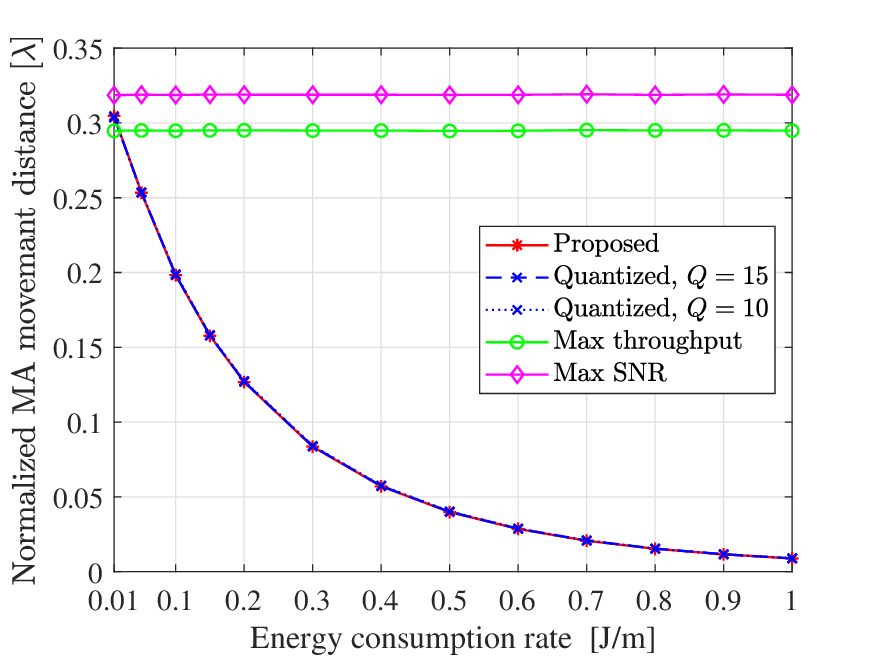}}
	\caption{Energy efficiency and normalized MA moving distance versus energy consumption rate.}
	\label{sg_E_MA}
\end{figure}
Fig. \ref{sg_E_MA} investigates the impact of energy consumption rate $\bar E$ on the energy efficiencies and normalized MA moving distances of various schemes. Generally, a higher energy consumption rate indicates a greater cost for MA movement. Therefore, the energy efficiencies of \textit{Max throughput} and \textit{Max SNR} schemes continuously decrease as $\bar E$ increases and drop below that of the \textit{FPA} scheme when $\bar E > 0.2$ J/m. However, their MA moving distances remain unchanged. This indicates that these two schemes are not adaptive to different stepper motor specifications, i.e., energy consumption rates. In contrast, the \textit{Proposed} and \textit{Quantized} schemes exhibit a decreasing MA moving distance as $\bar E$ increases, eventually approaching zero for large $\bar E$, which leads to a stable energy efficiency converging to that of the \textit{FPA} scheme. This is because when the cost of MA movement is too high, maintaining the initial MA position becomes the optimal choice for maximizing energy efficiency. Thus, in practical applications, it is recommended to use a low-power stepper motor to control the movement of a lightweight MA element at the terminal to achieve high energy efficiency. Moreover, as shown in Fig. \ref{sg_E_MA}\subref{E_MA_EE}, the energy efficiency of the \textit{Quantized} scheme is slightly lower than that of the \textit{Proposed} scheme. However, its MA moving distance is approximately equal to that of the \textit{Proposed} scheme in Fig. \ref{sg_E_MA}\subref{E_MA_mov_dis}. This indicates that the quantized visual AoDs do not introduce additional movement-related overhead to the \textit{Proposed} scheme in the single-user scenario. The energy efficiency loss of the \textit{Quantized} scheme is attributed to the degradation in communication performance due to inaccurate angle information. On the other hand, the \textit{Quantized} scheme performs nearly as well as the \textit{Proposed} scheme with perfect angle information, even with the virtual AoDs quantized at a resolution of just $Q=15$ or $Q=10$. This implies that the \textit{Proposed} scheme does not require high-precision angle estimation, which makes it appealing to implement in practical systems.

\begin{figure}[!t]
	\centering
	\includegraphics[width=1\linewidth]{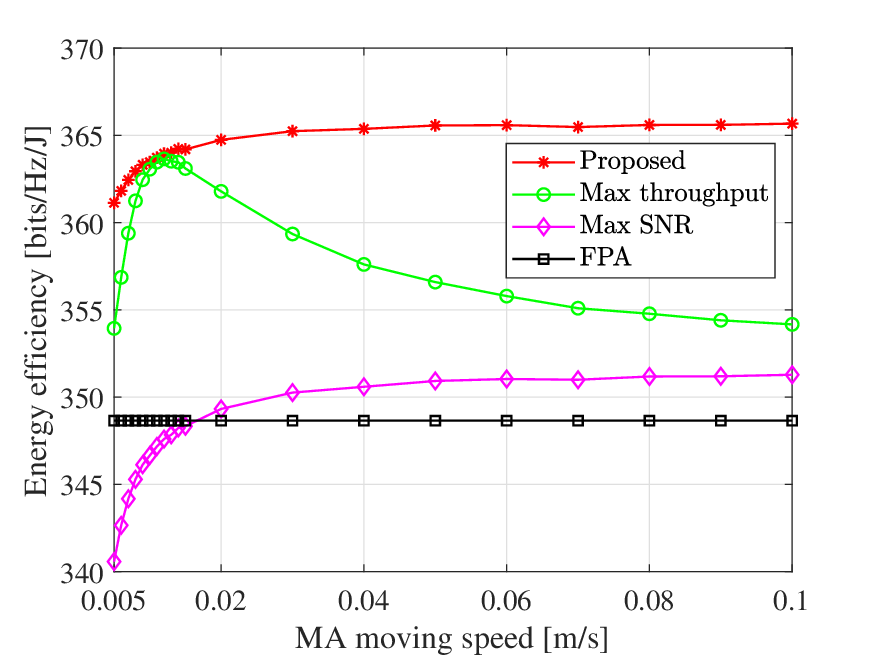}
	\caption{Energy efficiency versus MA moving speed.}
	\label{sg_v}
\end{figure}
\begin{figure}[!t]
	\centering
	\includegraphics[width=1\linewidth]{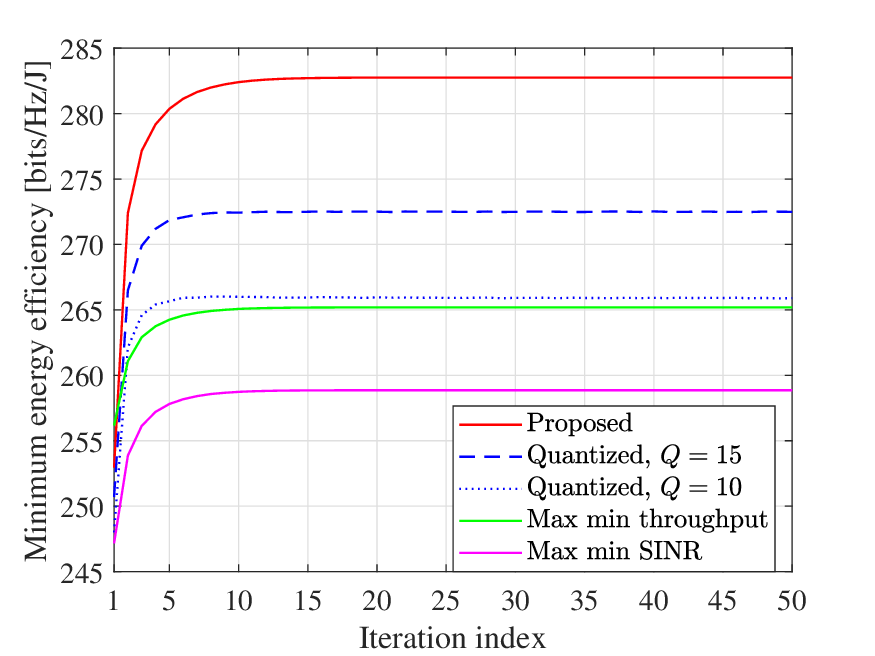}
	\caption{Convergence behaviors of Algorithm \ref{alg2} and benchmark schemes.}
	\label{conv}
\end{figure}
Fig. \ref{sg_v} investigates the energy efficiencies of different schemes under the scenario where the MA movement is severely constrained due to extremely low MA moving speed. In Fig. \ref{sg_v}, the MA moving speed $v$ is constrained between 0.005 m/s and 0.1 m/s, with a moving region length of $0.5\lambda$. To provide a clear comparison among schemes with different objectives, the \textit{Upper bound} and \textit{Quantized} schemes are omitted. We can see that the energy efficiency of the \textit{Max SNR} scheme improves with increasing $v$, which achieves performance close to the \textit{FPA} scheme at $v = 0.015$ m/s and subsequently surpasses it. This is because, in the \textit{Max SNR} scheme, a higher $v$ allows the MA to cover the same distance in less time, thereby leaving more time for data transmission. Moreover, the \textit{Max throughput} scheme exhibits a non-monotonic trend, with its energy efficiency initially increasing and then decreasing, and the peak performance approaching that of the \textit{Proposed} scheme. The reason is that, for low values of $v$, the performance gain of the \textit{Max throughput} scheme is attributed to the ability to explore a larger moving region with increasing $v$. However, as $v$ continues to increase, the energy overhead of MA movement outweighs the gains, which eventually degrades the overall performance. This also highlights the superior robustness of the \textit{Proposed} scheme across varying values of $v$.
\subsection{Multi-User System}
Next, we consider the multi-user scenario, i.e., $K>1$. We compare the minimum energy efficiency over users by the proposed algorithm in Algorithm \ref{alg2}, which is termed as \textit{Proposed}, with four benchmark schemes defined as \textit{Quantized}, \textit{Max min throughput}, \textit{Max min SINR}, and \textit{FPA}, which are similar to the benchmark schemes in Section \ref{simu_sg}.

Fig. \ref{conv} evaluates the convergence of Algorithm \ref{alg2} and compares it with benchmark schemes. As observed, the minimum energy efficiencies of all schemes increase with the iteration index and demonstrate rapid convergence, reaching stability within 15 iterations. Besides, the proposed iterative algorithm, i.e., Algorithm \ref{alg2}, results in an increase in the minimum energy efficiency from 252.95 bits/Hz/J to 282.76 bits/Hz/J, offering superior performance compared to benchmark schemes.

\begin{figure}[!t]
	\centering
	\subfloat[]{\label{v_EE}\includegraphics[width=1\columnwidth]{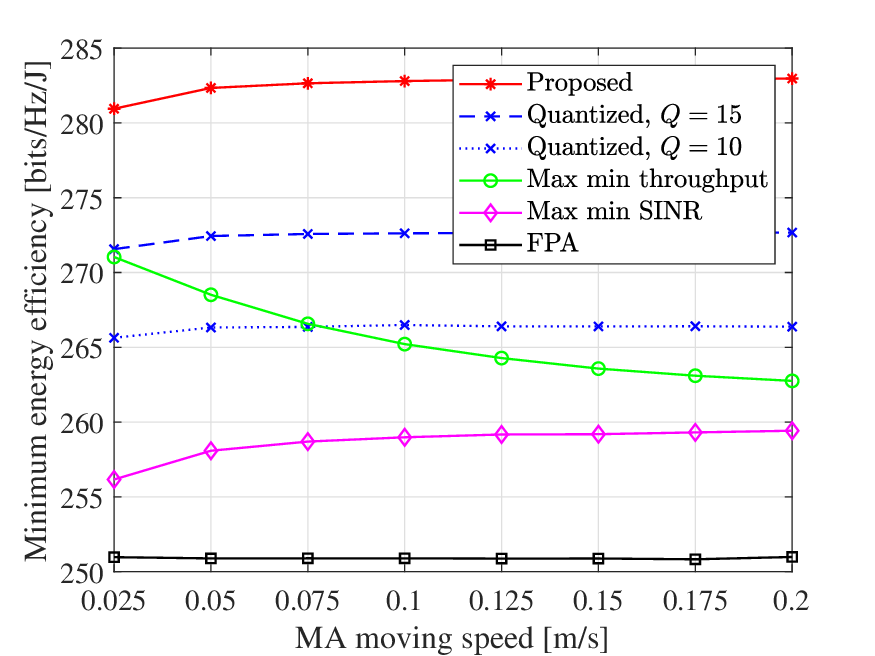}}\\
	\subfloat[]{\label{v_mov_dis}\includegraphics[width=1\columnwidth]{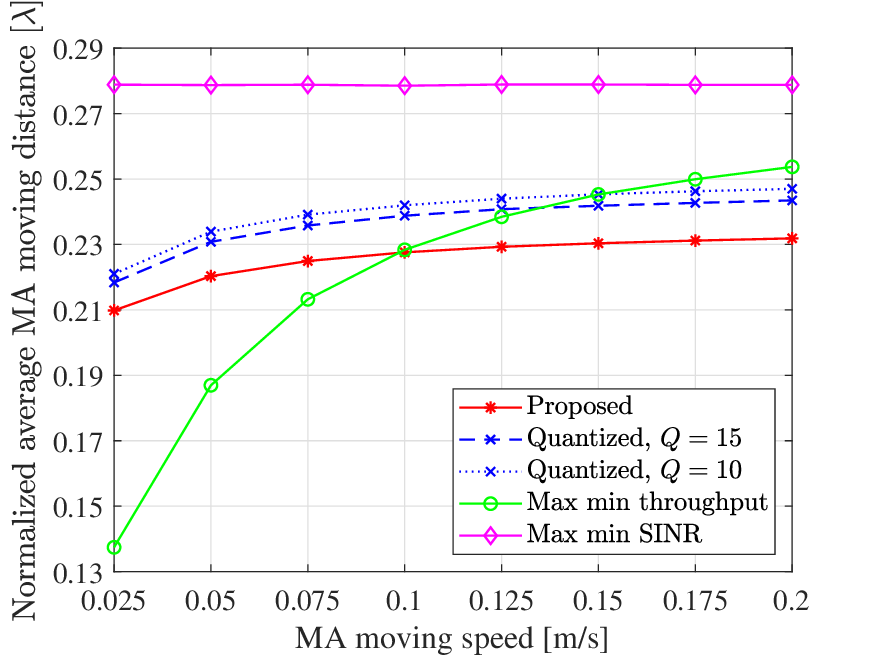}}
	\caption{Minimum energy efficiency and normalized average MA moving distance versus MA moving speed.}
	\label{mu_v}
\end{figure}
\begin{figure}[!t]
	\centering
	\subfloat[]{\label{T_EE}\includegraphics[width=1\columnwidth]{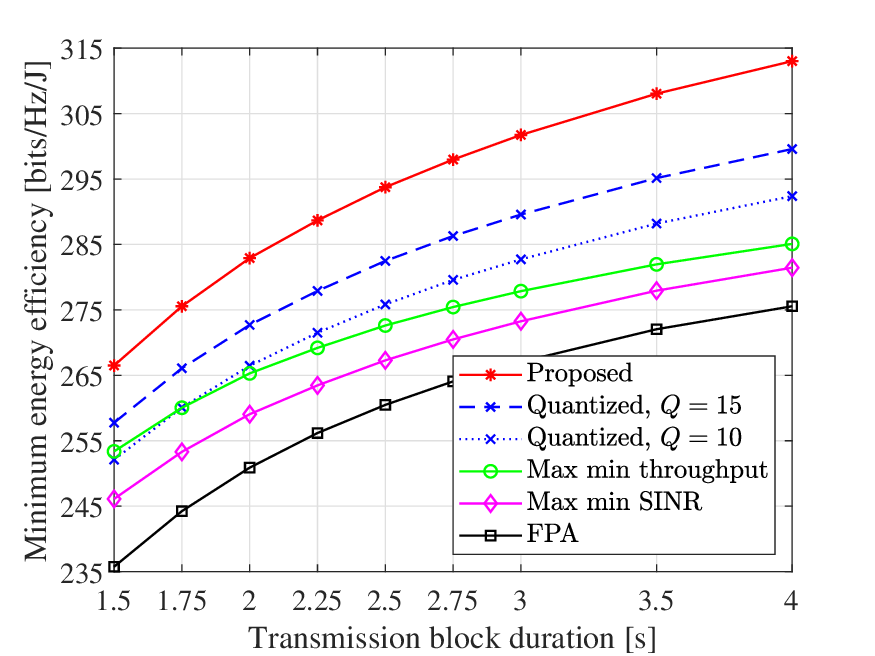}}\\
	\subfloat[]{\label{T_mov_dis}\includegraphics[width=1\columnwidth]{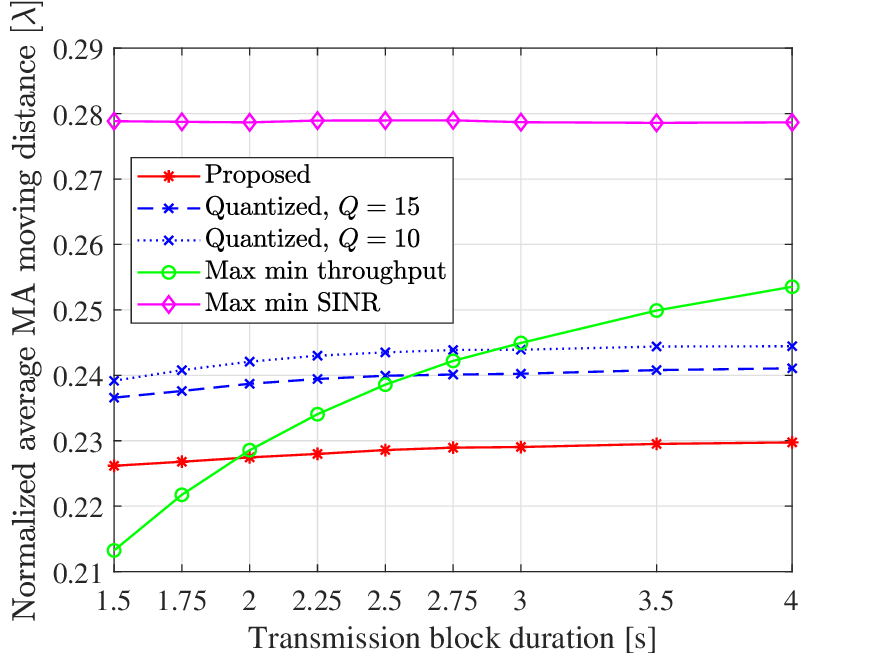}}
	\caption{Minimum energy efficiency and normalized average MA moving distance versus transmission block duration.}
	\label{mu_T}
\end{figure}
Fig. \ref{mu_v} illustrates the minimum energy efficiency and normalized average MA moving distance, i.e., ${\frac{1}{K}\sum\nolimits_{k = 1}^K {{{\left| {{x_k} - x_k^0} \right|}}/{\lambda }} }$, versus the MA moving speed $v$. As $v$ increases, the minimum energy efficiencies of the \textit{Proposed}, \textit{Quantized}, and \textit{Max min SINR} schemes increase, whereas that of the \textit{Max min throughput} decreases. The reason is as follows. For the \textit{Max min SINR} scheme, the average MA moving distance remains unaffected by variations in $v$ since its sole objective is to maximize the receive SINR of the signal from each user. As a result, an increase in $v$ reduces the MA moving delay, thereby allowing more time for communication within a given transmission block duration. For the \textit{Max min throughput} scheme, a higher $v$ allows the MA to move farther within a limited duration to explore more favorable channel conditions, thereby maximizing throughput. Nevertheless, the \textit{Max min throughput} scheme does not account for the energy consumption associated with MA movement, leading to a significant increase in the average MA moving distance as $v$ increases and, consequently, a decline in energy efficiency. The \textit{Proposed} and \textit{Quantized} schemes, which jointly consider both the time and energy consumptions of MA movement, effectively regulate the increase in MA moving distance, thus achieving superior energy efficiency as $v$ increases. Nonetheless, high-speed MA movement imposes stringent requirements on the accuracy of the antenna positioning module, which should be carefully considered as a trade-off in practical systems.

Fig. \ref{mu_T} depicts the minimum energy efficiencies and normalized average MA moving distances of various schemes for different transmission block durations $T$. It can be observed in Fig. \ref{mu_T}\subref{T_EE} that the performance of all schemes improves as $T$ increases, and the \textit{Proposed} scheme consistently achieves the best performance. Moreover, the \textit{Quantized} scheme shows relatively poorer performance in the multi-user system compared to the single-user system presented in Section \ref{simu_sg}. This is because the difference between the quantized and actual visual AoDs affects the antenna position optimization for each user, which increases the average MA moving distance compared to the \textit{Proposed} scheme (see Fig. \ref{mu_T}\subref{T_mov_dis}), thereby incurring additional movement-related overhead. It is worth noting that by merely increasing the quantization resolution from $Q=10$ to $Q=15$, the performance improves from 93.41\% to 95.71\% of that achieved with perfect angle information. Furthermore, as $T$ increases, the proportion of time delay associated with MA movement within the entire transmission block duration decreases, which makes the performance advantage of the \textit{Proposed} scheme over the \textit{FPA} scheme more significant.
\section{Conclusion}\label{6}
This paper investigated energy efficiency maximization in an MA-aided multi-user uplink communication system, where each user is equipped with a single MA capable of linear movement. We modeled the energy consumption of the stepper motor in the antenna positioning module as a function of the MA's initial and optimized positions, and proposed corresponding algorithms to jointly design the receive combining matrix of the BS, as well as the transmit power and MA position of each user. We first considered the single-user scenario. An optimization algorithm based on the 1D exhaustive search was proposed, and the upper bound on the energy efficiency was derived. Then, for the multi-user scenario, we developed an iterative algorithm to efficiently maximize the minimum energy efficiency among all users. Simulation results demonstrated the effectiveness of the proposed scheme in enhancing the users' energy efficiencies and provided valuable insights for practical applications.
\section*{Acknowledgments}
The calculations were supported by the High-Performance Computing Platform of Peking University.
\appendices
\section{Proof of Proposition \ref{proposition2}}\label{appendixB}
For the given transmit power $p^\star$, the user's energy efficiency can be written as
\begin{align}
	EE & = \frac{{\left( {T - \frac{{\left| {x - {x^0}} \right|}}{v}} \right){{\log }_2}\left( {1 + \frac{{p^\star\left\| {\mathbf{h}\left( x \right)} \right\|_2^2}}{{{\sigma ^2}}}} \right)}}{{\bar E\left| {x - {x^0}} \right| + \frac{p^\star}{\eta_\mathrm{c}}\left( {T - \frac{{\left| {x - {x^0}} \right|}}{v}} \right)}}  \nonumber\\
	&=\frac{{{{\log }_2}\left( {1 + \frac{{p^\star\left\| {\mathbf{h}\left( x \right)} \right\|_2^2}}{{{\sigma ^2}}}} \right)}}{{\frac{{\bar E\left| {x - {x^0}} \right|}}{{T - \frac{{\left| {x - {x^0}} \right|}}{v}}} + \frac{p^\star}{\eta_\mathrm{c}}}} \nonumber\\
	&\mathop \le \limits^{\left( {{b_1}} \right)} \frac{\eta_\mathrm{c}}{{p^\star}} {{{\log }_2}\left( {1 + \frac{{p^\star\left\| {\mathbf{h}\left( x \right)} \right\|_2^2}}{{{\sigma ^2}}}} \right)} \nonumber\\
	&\mathop \le \limits^{\left( {{b_2}} \right)} \frac{\eta_\mathrm{c}}{{p^\star}}{{{\log }_2}\left( {1 + \frac{{p^\star\left\| {\mathbf{h}\left( {\bar x} \right)} \right\|_2^2}}{{{\sigma ^2}}}} \right)} \triangleq E{E^\mathrm{ub}},
\end{align}
where the inequality marked by $\left( {{b_1}} \right)$ holds because ${{{\bar E\left| {x - {x^0}} \right|}}/\left( {{T - {{\left| {x - {x^0}} \right|}}/{v}}}\right) } \ge 0$, and the equality can be achieved when $x=x^0$. The inequality marked by $\left( {{b_2}} \right)$ holds because ${\left\| {\mathbf{h}\left( \bar x \right)} \right\|_2^2}$ is the maximum channel gain, and the equality can be achieved when $x=\bar x$. Hence, when $x^0=\bar x$, the upper bound on the user's energy efficiency, $EE_\mathrm{ub}$, can be achieved. This thus completes the proof.
\section{Construction of $\varepsilon_{jk}$}\label{appendixC}
Based on the expression of ${{\mathrm{d}{h_{jk}}\left( {x_j^i} \right)}}/{{\mathrm{d}{x_j}}}$ in \eqref{1order}, we have
\begin{small}
\begin{align}
	&\frac{{{\mathrm{d}^2}{h_{jk}}\left( {x_j^i} \right)}}{{\mathrm{d}x_j^2}} \nonumber\\
	&= \sum\limits_{a = 1}^{{L_k}-1} \sum\limits_{b = a+1}^{{L_k}}  - \frac{{8{\pi ^2}\left| {{m_{jk,ab}}} \right|\vartheta _{k,ab}^2}}{{{\lambda ^2}}}\cos \left( {\frac{{2\pi}}{\lambda }x_j^i{\vartheta _{k,ab}} + \angle {m_{jk,ab}}} \right)  ,
\end{align}
\end{small}
and can select $\varepsilon_{jk}$ as
\begin{equation}\label{varkj}
	\varepsilon_{jk}=\sum\limits_{a = 1}^{{L_k}-1} {\sum\limits_{b = a+1}^{{L_k}} { \frac{{8{\pi ^2}\left| {{m_{jk,ab}}} \right|\vartheta _{k,ab}^2}}{{{\lambda ^2}}}} }.
\end{equation}

\end{document}